\documentclass{aa}
\usepackage{graphicx}
\usepackage{setspace}
\usepackage{amsmath,amssymb,amsfonts}
\usepackage{txfonts}
\usepackage{epstopdf}
\usepackage{color}
\usepackage{natbib}
\begin{document}

\title{H {\sc i} Kinematics of the Large Magellanic Cloud revisited : Evidence of possible infall and outflow}

\author{Indu, G.\inst{1,2}, Annapurni Subramaniam\inst{1}}
\institute{Indian Institute of Astrophysics, Koramangala II Block, Bangalore-560034, India\\
		Pondicherry University, R. Venkataraman Nagar, Kalapet, Pondicherry-605014, India\\
           \email{indu@iiap.res.in, purni@iiap.res.in}}
\date{Received, accepted}
\abstract
{}
{The neutral atomic Hydrogen (H {\sc i}) kinematics of the Large Magellanic Cloud (LMC) is revisited in light of two new proper motion estimates.}
{We analysed the intensity weighted {H {\sc i}} velocity maps of the Australia Telescope Compact Array (ATCA)/Parkes and the Parkes Galactic all sky survey (GASS) data sets. We corrected the line-of-sight velocity field for the systemic, transverse, precession, and nutation motions of the disk using two recent proper motion estimates, and estimated the kinematic parameters of the {H {\sc i}} disk.}
{The value of position angle (PA) of kinematic major axis estimated using ATCA/Parkes data ($126^\circ\pm23^\circ$) is found to be similar to the recent estimate of the PA using stellar tracers. The effect of precession and nutation in the estimation of PA is found to be significant. The modelled {H {\sc i}} disk is found to be disturbed within 1.$^o$0 radius and beyond 2.$^o$9 radius. Using ATCA/Parkes data, most of the {H {\sc i}} gas in the LMC ($\sim 87.9\%$ of the data points) is found to be located in the disk. We detected $12.1\%$ of the data points as kinematic outliers. A significant part of type 1 as well as slow type 2 {H {\sc i}} gas is identified with Arm E. We identified the well-known Arm S, Arm W, Arm B and a new stream, Outer Arm, as part of fast type 2 outlier component. The GASS data analysis brings out the velocity details of the Magellanic Bridge (MB) and its connection to the LMC disk. We find that the Arm B and the Outer Arm  are connected to the MB. We detect high velocity gas in the western disk of the LMC and the south-west and southern parts of the MB.}
{We proposed two models (in-plane and out-of-plane) to explain the outlier gas. We suggest that the Arm B could be an infall feature, originating from the inner MB. The Arm E could be an outflow feature. We suggest possible outflows from the western LMC disk and south and south-western MB, which could be due to ram pressure. The velocity pattern observed in the MB suggests that it is being sheared. We suggest that the various outliers identified in this study may be caused by a combination of tidal effects and hydrodynamical effect due to the motion of the LMC in the Milky Way (MW) halo.}

\keywords{galaxies: Magellanic Clouds;
galaxies: evolution;
galaxies: kinematics and dynamics 
}
\authorrunning{Indu \& Subramaniam}
\titlerunning{H {\sc i} Kinematics of the LMC}
\maketitle

\section{Introduction}
\hspace{4ex}Our nearest neighbours, the Magellanic Clouds continue to be the two most studied objects in extra galactic astronomy. The morphology of the Large Magellanic Cloud (LMC) in optical wavelength is dominated by regions of strong star formation and dust absorption. The distribution of the neutral atomic Hydrogen (H {\sc i}) in the LMC is dictated by filaments combined with shells and holes. The structural parameters of gaseous as well as stellar distribution are estimated by various studies. The summarised viewing angles are, an inclination in the range $25^\circ - 35^\circ$ and the position angle (PA) of line of nodes (LON) in the range $120^\circ - 160^\circ$ (\citealt{2009IAUS..256...81V}, \citealt{2010A&A...520A..24S}) for the stellar distribution, estimated with kinematic and photometric analyses. Recent estimates of the stellar kinematic major axis are, $131^\circ$, using carbon stars, with di/dt $\sim$ 0.0 \citep{2007ApJ...656L..61O} and $142^\circ\pm5^\circ$, using red super giants, with di/dt $-184^{\circ}/Gyr$ \citep{2011ApJ...737...29O} both estimated with the new proper motion values of \citet{2008AJ....135.1024P} and \citet{2006ApJ...638..772K}, where di/dt is the rate of change of inclination that is a measure of precession and nutation of the LMC disk. The PA of kinematic major axis for the gaseous distribution, estimated using the old proper motion \citep{1994AJ....107.1333J}, without a di/dt correction is about $168^\circ$ \citep{1998ApJ...503..674K}, which clearly indicates the discrepancy between the kinematic parameters estimated using gaseous and stellar tracers. 
This complicates the study and analysis of internal kinematics and the underlying gravitational potential. Thus, a re-estimate of H {\sc i} kinematics of the LMC disk using the recent proper motion estimates is necessary. It has been recently realised that including the precession and nutation of the disk can change the estimated kinematics. Various features in the gas have been found that are connected to the Magellanic Bridge (MB), the Small Magellanic Cloud (SMC), and the Leading Arm (LA), but their formation and the details of the gas flow in these features are not clearly understood. \citet{2003MNRAS.339...87S} identified various arm-like H {\sc i} filaments (Arm E, B, S \& W) in the H {\sc i} distribution of the LMC.
Kinematically different components in the gas and stars have been identified by various studies (\citealt{2011ApJ...737...29O}, \citealt{2012ApJ...753..123C}) and the reasons for the formation of these outliers are only suggestive. \citet{2011ApJ...737...29O} found a kinematically distinct population of stars that are possibly counter-rotating and accreted from the SMC. These stars constitute around $5\%$ of the total sample and these are associated with Arm E \& B. Hence \citet{2011ApJ...737...29O} proposed that Arm B and E may be signatures of gas accretion. They also suggested a possibility that arms could be counter-rotating. On the other hand if there is gas accretion present in the LMC, details on the possible direction of accretion, locations of accretion in the disk, etc. are not known. \citet{2008ApJ...679..432N} found that a part of Arm B is connected to the Magellanic Stream (MS). Later \citet{2013ApJ...772..110F} and \citet{2013ApJ...772..111R} measured the metallicity of the MS in the direction of this filament and found it to be similar to the LMC metallicity. This validates the connection of this filament and its origin in the LMC as proposed by \citet{2008ApJ...679..432N}. \citet{2008ApJ...679..432N} connected the LA feature with the Arm E physically. They suggested that the LA and the stream connected to the MS originate from the H {\sc i} over-density region near 30 Doradus and gas leaves the LMC through both these streams. Thus, it is not quite clear whether there is gas infall or gas expulsion through these arms. We have performed a detailed analysis using H {\sc i} data in an attempt to address the above issues.

The recent proper motion estimates of \citealt{2008AJ....135.1024P}, \citealt{2006ApJ...638..772K}, \citealt{2006ApJ...652.1213K} and \citealt{2013ApJ...764..161K} are very different from the older values and the estimation of the H {\sc i} velocity field in the outer disk is affected by this change. Hence it is necessary to compute the H {\sc i} velocity field in the LMC using the new proper motion values.  
In this study, we computed the kinematic PA and the circular velocity curve using the data of \citet{2003ApJS..148..473K}, after correcting it for the transverse motion using the new proper motion estimates. 
The unresolved puzzles are the presence and origin of kinematical differences between stellar and gaseous distributions and the identification of the location, amplitude and origin of the non-equilibrium features and their kinematics.

It is obvious that the structure, kinematics and the star formation history of the LMC is modified by its interaction with the Milky Way (MW) and the SMC, even according to the first passage scenario. \citet{2001AJ....122.1827V} found that the LMC is elongated in the general direction of the Galactic centre and is elongated perpendicular to the MS and velocity vector of the LMC centre of mass. He suggested that the elongation of the LMC has been induced by the tidal force of the MW. According to the predictions of the N body simulations by \citet{2000ApJ...532..922W}, the MW tidal torques can cause the precession and nutation of the LMC disk plane. \citet{2005MNRAS.363..509M} claimed the combined effect of gravity and ram pressure can account for the majority of the LMC's kinematical and morphological features. According to them, the tidal forces exerted by the gravitational potential of the Galaxy elongates the LMC disk, forcing a bar and creating a strong warp and diffuse stellar halo. The star formation in the central disk is dominated by the LMC bar. By applying new proper motion, \citet{2009IAUS..256..117M} found that the effect of tidal forces on the vertical stellar structure of the LMC are marginal and the tidal stripping is almost effectless. \citet{2009ApJ...703L..37S} and \citet{2007PASA...24...21B} suggested that the off-centred bar of the LMC could be the result of the collision with the SMC.

In general, the recent star formation in the Magellanic Clouds have been attributed to their mutual interaction and their hydrodynamic interaction with the Galaxy. \citet{2009MNRAS.399.2004M} analysed the effect of the ram pressure on the gas distribution and star formation in the LMC. They found a compression of the gas in the leading edge. They also found enhanced star formation activities in the leading border and eastern part of the disk (where the 30 Dor and most of the super-giant-shells (SGS) are located) in the last 30 Myr.  The study by \citet{2011A&A...535A.115I} on the recent star formation history of the Large and Small Magellanic Clouds (L\&SMC) shows that the recent star formation in both the Clouds is dictated by the perigalactic passage. In the first passage scenario, the LMC has probably had its closest approach about 200 Myr ago. This event is likely to have had a significant impact on both the Clouds, especially on the LMC. There is an enhanced star formation detected in the northern regions of the LMC disk around 200 Myrs ago and in the north-eastern part of the disk in the last 40 Myrs. This was explained as a combined effect of the gravitational/tidal interaction of the Galaxy and the effect of ram pressure due to the motion of the LMC in the Galactic halo towards north-east (NE).

\citet{2001AJ....122.1827V} calculated that the tidal force from the MW on the LMC is 17 times larger than that from the SMC. He suggested that the tidal effect of the SMC on the LMC is negligible compared with that of the MW. \citet{2012MNRAS.421.2109B} illustrated that the observed irregular morphology and internal kinematics of the Magellanic System (in gas and stars) are naturally explained by interactions between the LMC and the SMC, rather than gravitational interactions with the Galaxy. They also demonstrated that the off-centred, warped stellar bar of the LMC and its one-armed spiral, can be naturally explained by a recent direct collision with its lower mass companion, the SMC. A stellar counterpart of the H {\sc i} MB that was tidally stripped from the SMC, $\sim$200 Myr ago during a close encounter with the LMC, was discovered recently by \citet{2013ApJ...779..145N}.

There are numerous observations and analyses in the various impacts of the LMC-SMC-MW interactions. The gas distribution in the LMC might definitely bear the signatures of its tidal interaction with the SMC and the Galaxy and also the hydrodynamical interaction with the Galactic halo. A detailed study in this direction can reveal more about the LMC-SMC-MW interaction. In this study, we estimated the modelled H {\sc i} disk of the LMC and presented the kinematical properties of the H {\sc i} disk, and hence kinematic outliers are identified.  We studied the kinematics of the LMC to an extended region of about three to four times the area of the Australia Telescope Compact Array (ATCA)/Parkes data, which also includes the MB. We utilized the Parkes Galactic all sky survey (GASS) data set for this purpose. We used two recently published values of proper motion, to compare their effect on the estimation of the kinematic properties. We also demonstrate the effect of precession and nutation (caused by the MW tidal torques) on the estimated kinematic parameters as a function of radius. 
The paper is organised as follows. Section 2 describes the data and section 3 outlines our methodology with sub-section for various procedures. Section 4 describes the estimation of residuals and various outliers and section 5 is the GASS data analysis. The two scenarios to explain the origin of outliers are given in section 6. The results and discussion are presented in section 7, and the conclusions are summarised in section 8.

\section{Data}
\hspace{4ex}We make use of two different H {\sc i} velocity data sets for the analysis. They are ATCA/Parkes and GASS data sets, the details of which are given in the following sections. The ATCA/Parkes data cover only the disk of the LMC and have the highest spatial resolution to study the LMC kinematics in detail. The GASS provides all sky survey data, which is ideal for analysing the kinematics to a larger spatial extent.
\subsection{ATCA/Parkes}
\hspace{4ex}In this study, we used the published H {\sc i} velocity data of the LMC from \citet{2003ApJS..148..473K} observed with ATCA and Parkes telescope. The spatial resolution of the combined data set is $1^\prime$ and its velocity resolution is 1.65 km s$^{-1}$. The area covered is around $13^{\circ}$ $\times$ $14^{\circ}$. The spatial resolution of the ATCA/Parkes data is helpful to analyse the LMC disk kinematics in detail. For an extended study we used an all sky survey data set.
\subsection{GASS} 
\hspace{4ex}The GASS data (\citealt{2009ApJS..181..398M} and \citealt{2010A&A...521A..17K}) is a survey of Galactic H {\sc i} emission in the southern sky observed with the Parkes 64 m radio telescope. The GASS is known to be the H {\sc i} all sky survey with  the highest spatial resolution, sensitivity, and accuracy currently available for the southern sky. We used the stray radiation corrected second data release. The stray radiation correction was done by convolving the all sky response of the Parkes antenna with the brightness temperature distribution from the Leiden/Argentine/Bonn (LAB) survey. The spatial resolution is $4^\prime.8$ and the velocity resolution is of the order of 0.8 km s$^{-1}$. The sky covered is with $\delta \le 1^\circ$. GASS data set has 1137 velocity frames from which the velocity frames for the LMC and the MB are selected. For GASS we have an area coverage of $20^{\circ}$ x $20^{\circ}$.  

It is important to note that we used the first moment map or the intensity weighted mean velocity map  of both the  data sets for the analysis. Since we use  the mean velocity for each pixel, the data is free from the multi-peak velocity profiles. According to \citet{1998ApJ...503..674K} even though such a map is an imperfect way to analyse complex structure, it represents the best estimate of the mean velocity field that can be derived from the data set. Hence our analysis may only bring out average/bulk kinematic features that leave a major signature in the mean field. The method is insensitive to features with much less/insignificant contribution to the H {\sc i} flux.
Here we utilized the extensive sky coverage of the GASS survey to trace the H {\sc i} features in and around the LMC and the MB.
 
\section{Methodology}
\subsection{Transverse motion correction}
\hspace{4ex}The observed line-of-sight velocity is required to be corrected for the three-dimensional space motion, precession, and nutation of the system. For the LMC, we used the recent proper motion estimates by \citet{2008AJ....135.1024P} and \citet{2013ApJ...764..161K}. The line-of-sight velocity field of a system at a point with an angular distance $\rho$ from the centre and PA $\phi$ is given by 
equation 24 of \citet{2002AJ....124.2639V},
\begin{multline*}
V_{los}(\rho,\phi) = V_{sys} \cos \rho + V_t \sin \rho \cos (\phi - \theta_t)  \\ + D_0 (di/dt) \sin \rho \sin (\phi - \theta) \\ - s V(R) f \sin i \cos(\phi - \theta).
\end{multline*}
In the above equation, $V_{sys}$ is the systemic velocity, $V_t$ is the transverse velocity, and $\theta_t$ is the PA of the transverse velocity vector in the sky. The first and second terms correspond to the systemic and transverse motion. The angle of inclination of the plane of the LMC is denoted by $i$ and the PA of kinematic LON by $\theta$. The distance to the galaxy is $D_0$. The third term on the right-hand side represents the precession and nutation of the disk of the galaxy. The last term represents the internal rotational motion of the galaxy with $V(R)$ as in-plane rotational velocity. The quantity $s$ is called the spin sign of rotation and $f$ is a function of $\rho$, $\phi$, $\theta$ and $i$. By simplifying  and rearranging the above relation, we obtain the velocity field of the LMC as,
\begin{multline}
 V_{mod}(\rho,\phi) = V_{los}(\rho,\phi) - V_{sys} \cos \rho - V_t \sin \rho \cos(\phi - \theta_t) \\- D_0 (di/dt) \sin \rho \sin(\phi - \theta).
\end{multline}
We keep the effect of inclination and do not do a de-projection. 
We used the H {\sc i} kinematical centre, $\alpha_0 = 5^h 17.6^m$, $\delta_0 = -69^\circ 02^\prime$  and $V_{sys} = 279$ km s$^{-1}$ estimated by \citet{1998ApJ...503..674K}. According to \citet{2001AJ....122.1807V}, the choice of centre can be arbitrary to some extent since it does not effect the modelling significantly.
Here we have used the di/dt values $-184^\circ\pm80^\circ/Gyr$ \citep{2011ApJ...737...29O} and $-103^\circ\pm61^\circ/Gyr$ \citep{2002AJ....124.2639V} and compared the estimates. We used the value of $\theta$ to be $138^\circ$, our initial estimate. The proper motion values $\mu_W$ \& $\mu_N$, (\citealt{2008AJ....135.1024P} and \citealt{2013ApJ...764..161K}) are converted to linear velocities, $V_x$ \& $V_y$ using equation 28 of \citet{2002AJ....124.2639V}. We estimated the transverse velocity $V_t$ as the resultant of $V_x$ \& $V_y$ and also estimated its direction $\theta_t$. 
The values are tabulated in Table 1 with the first row taking values of \citet{2008AJ....135.1024P} and the second row taking the values of \citet{2013ApJ...764..161K}. \citet{2002AJ....124.2639V} using the mean of the proper motion estimates available at the time of their study, quote the values of $V_x$ \& $V_y$ to be -399 km s$^{-1}$ and 80 km s$^{-1}$, respectively. Their estimate of $V_t$ is 406 km s$^{-1}$ and $\theta_t$ is $78^\circ.7$. \\ 
The map of the line-of-sight velocity corrected for systemic, transverse, precession, and nutation motion, $V_{mod}$,  estimated using ATCA/Parkes data is shown in figure 1. (We used di/dt = $-184^\circ/Gyr$, and $V_t$ = 474.1 km s$^{-1}$.) The equatorial coordinates ($\alpha$, $\delta$)  are converted to Cartesian coordinates in kpc in a projected plane with respect to the LMC centre \citep{2001AJ....122.1807V}. This scheme is followed to estimate all the maps in this paper. While analysing ATCA data for the first time, \citet{1998ApJ...503..674K} did the transverse motion correction of the observed velocity field with the proper motion estimates of \citet{1994AJ....107.1333J} with a transverse velocity 286 km s$^{-1}$. Further the analysis is done with an approximation of di/dt = 0, not accounting for the precession and nutation of the disk. The precession and nutation plays an inevitable role in kinematical study because it contributes a velocity gradient perpendicular to the PA of LON. Figure 3 of \citet{1998ApJ...503..674K} shows the velocity map. The PA of kinematic major axis can be identified as almost vertical in the north-south (NS) direction  with an average value of $168^\circ$. One can vividly see the difference in the velocity distribution and the rotational shift in the PA of the apparent kinematic major axis in our figure 1, which is corrected for transverse motion with the new proper motion estimates and also corrected for the precession and nutation of the LMC disk, as described above. Thus, there is an apparent change in the kinematic major axis of the H {\sc i} disk, and hence there is a requirement to re-estimate the parameters of the H {\sc i} disk, such as the PA and circular velocity as a function of radius. We performed this task
and the details are discussed in the following sections. 
\begin{figure}
   \resizebox{\hsize}{!}{\includegraphics{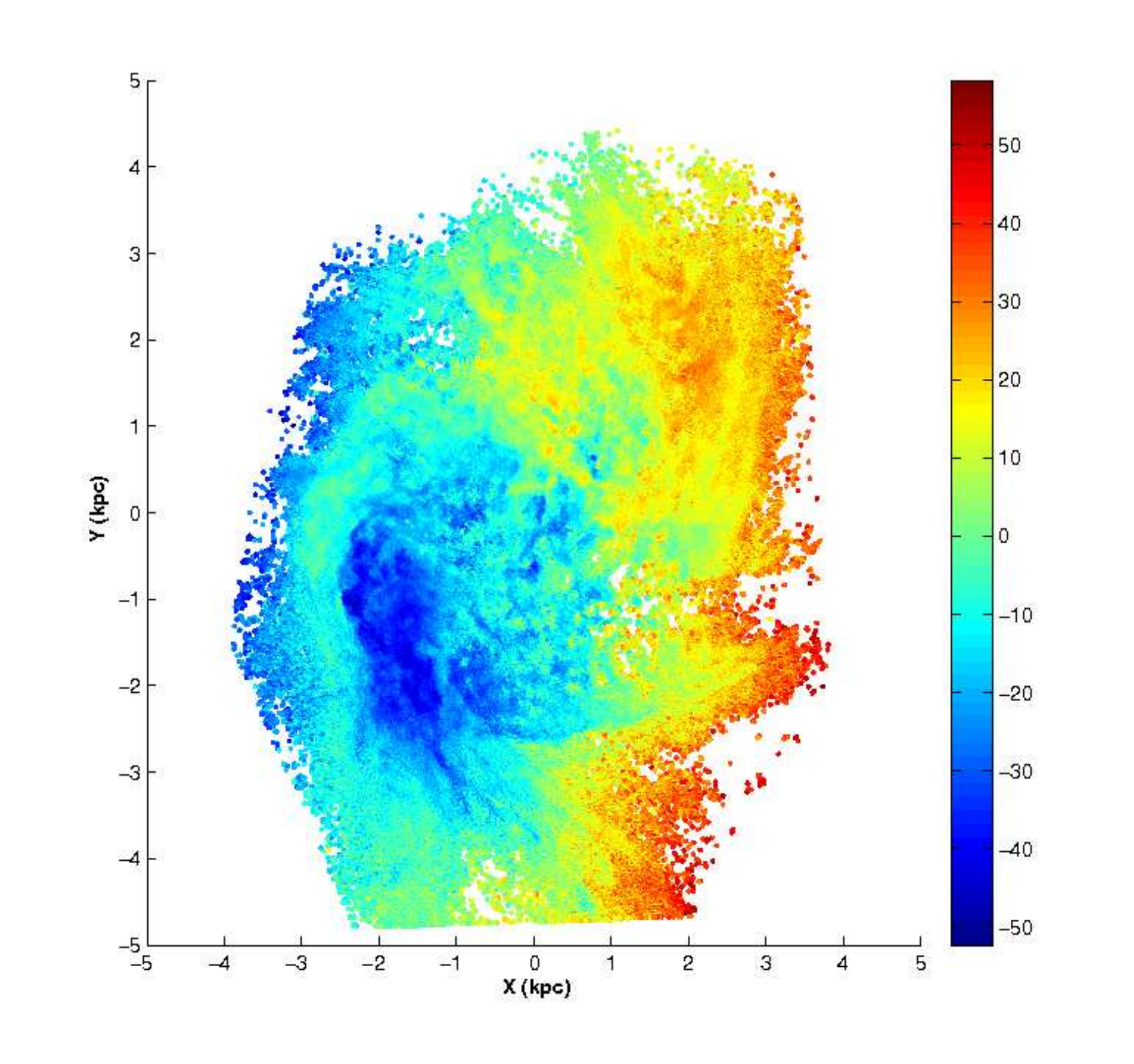}}
   \caption{Intensity weighted mean velocity map of the LMC H {\sc i} using ATCA/Parkes data, after correcting for transverse, systemic, precession, and nutation motions. Colour coding is according to the variation in $V_{mod}$ in km s$^{-1}$.}
    \end{figure}
\begin{table*}
      \caption[]{Transverse velocity estimates using two proper motion measurements.}
         \label{Table:1}
	\centering
	\begin{tabular}{c | c | c | c | c | c}
	\hline\hline
	$\mu_W$(mas/year) & $\mu_N$(mas/year) & $V_x$(km s$^{-1}$) & $V_y$(km s$^{-1}$) & $V_t$(km s$^{-1}$) & $\theta_t$($^\circ$) \\
	\hline
 	$-1.96\pm0.04^1$ & $0.440\pm0.04^1$ & -462.6 & 103.8 & $474.1\pm9.4$ & $77.4\pm1.2$ \\
	\hline
 	$-1.91\pm0.02^2$ & $0.229\pm0.047^2$ & -450.8 & 54.0 & $454.0\pm4.9$ & $83.2\pm1.4$ \\
	\hline
\end{tabular}
	\\{\it \tiny References : (1) \citet{2008AJ....135.1024P}; (2) \citet{2013ApJ...764..161K}}
   \end{table*}
\subsection{Annular ring analysis}
\hspace{4ex}We used the following method to study the H {\sc i} disk of the LMC. The intensity weighted mean velocity field for the ATCA/Parkes data corrected for transverse, systemic, precession, and nutation motions  is estimated as mentioned above. The disk of the LMC is divided into annular rings of width $0^\circ.2$, according to the available resolution of the data set. In each annular ring, we fitted a sine curve to the velocity field, the PA of maxima of which will be the PA of the kinematic major axis. We used the following equation,
\begin{equation}
V_{fit}(\rho,\phi) = V\, \sin(\phi - \phi_0) + \delta V_{sys}
\end{equation}  
where $V$ represents the amplitude of the sine curve, which is expected to increase as a function of radius. The variable $\phi$ represents the PA measured from north towards east and $\phi_0$ is the zero point of the sine curve (the point where $V_{fit}$ = $\delta V_{sys}$). The PA of kinematic major axis is derived as $\phi_0$ $\pm$ $90^\circ$. We estimated the parameters for 29 annular rings covering a radius of $5^\circ.7$ of the LMC.  Figure 2 demonstrates the fitted sine curves for annular rings at mean radii of $0^\circ.1$, $0^\circ.9$, $1^\circ.9$, $2^\circ.9$, $3^\circ.9$ \& $4^\circ.9$. From the first fit, we removed points that deviate from the model by $\ge$ 12 km s$^{-1}$, (the choice of this cut-off is explained in section 4) and the fit was repeated to make a refined model. The green points are the removed outliers, and only the black points are used for the refit. The refit thus excludes all possible outliers and the estimated parameters are likely to be as close as possible to the true disk parameters. One can see the amplitude of the sine curve increases with increasing radius. It can be seen that in the panel f for $4^\circ.9$ radius the data points are sparse. This happens in the outer most seven annular rings from a radius of $4^\circ.5$ to $5^\circ.7$. After a radius of $4^\circ.3$, the azimuthal variation of velocity along the annular rings does not provide a continuous sine curve due to poor data coverage.\\  
\begin{figure}
   \resizebox{\hsize}{!}{\includegraphics{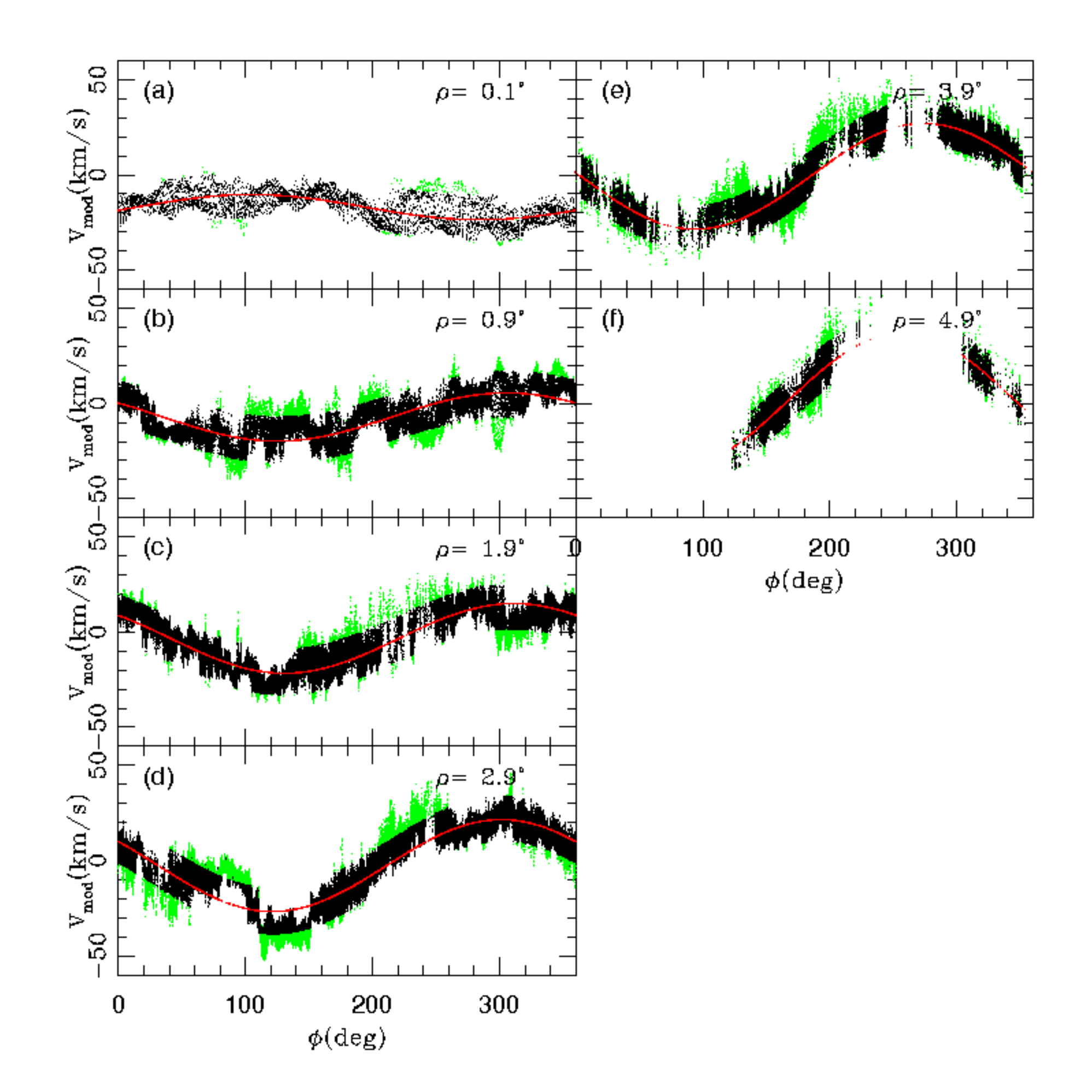}}
   \caption{Azimuthal variation of $V_{mod}$ and the fitted sine curves for annular rings. Panel (a) at mean radii of $0^\circ.1$, (b) $0^\circ.9$, (c) $1^\circ.9$, (d) $2^\circ.9$, (e) $3^\circ.9$, \& (f) $4^\circ.9$. The green points are the removed outliers, and only the black points are used for the refit.}
    \end{figure} 
Panel a of figure 3 shows the variation of the estimated PA as a function of radius with and without the precession and nutation correction. The curves for two different estimates of di/dt and $V_t$ are shown. The overall effect is that as the magnitude of di/dt increases the estimated PA decreases. The variation of PA across the radius is only mild, except in the central region. 
The mean values of PA estimated for different radial extends are tabulated in Table 2. The first row lists the estimate for the whole radius range till $5^\circ.1$ after which the fit error is more than $\sim 11\%$. The second row lists the estimated PA values omitting the central $0^\circ.5$ since the central region shows a twist in the PA with values $>$ $180^\circ$. The PA values decrease after a radius of $4^\circ$. Hence omitting the values after $4^\circ$ the mean PA is tabulated in the third row. 
The mean values of PA tabulated in the fourth and fifth columns are estimated with different $V_t$ values, but the same $di/dt$. There is no significant change in the value of mean PA.    
The last column lists the calculated standard deviation for the PA estimates in the fourth column. (The standard deviation estimates for the corresponding PA estimates in other columns are found to be comparable with these values.) Previous estimates for the stellar kinematic major axis are, $131^\circ$, using carbon stars, with di/dt $\sim$ 0.0 \citep{2007ApJ...656L..61O} and $142^\circ\pm5^\circ$, using red super giants, with di/dt $-184^{\circ}/Gyr$ \citep{2011ApJ...737...29O}, which are estimated with the new proper motion. These values are comparable to our values of the kinematic major axis within the estimated errors. Hence, our results suggest the kinematic major axis of gaseous and stellar components of the LMC are similar. The panel b of figure 3 shows the radial variation of circular velocity inferred from annular ring analysis.  The solid curves shown are not the conventional rotation curves since we have not performed inclination correction, and the velocity distributions include the effect of inclination. The velocity varies almost linearly up to a radius of $2^\circ.9$ and then shows a dip and again rises. 
Also, the velocity increases directly with the magnitude of di/dt. Both panels a and b of Figure 3 also demonstrate that the kinematic parameters of the H {\sc i} disk estimated using the proper motion values of \citet{2008AJ....135.1024P} and \citet{2013ApJ...764..161K} remains the same within errors. Hence, for further analysis, we used the proper motion estimates of \citet{2008AJ....135.1024P}. This also helps to compare the estimated values with the stellar kinematical parameters estimated by \citet{2011ApJ...737...29O}. The bold curve in orange shows the resultant curve from the refined model (after the first iteration, removing the outliers) the parameters estimated from which are used for further analysis.

\citet{2013A&A...552A.144S} in their Table 1 provided a summary of orientation measurements of the LMC disk plane. The table lists out the estimated inclination and PA values in literature, using various tracers. We estimated the weighted average of the inclination values in the table, which is $\sim 25^\circ$. The de-projected rotation curve  estimated, applying this inclination, is shown as a dotted curve in figure 3 panel b. 
The value of $V_{rot}$ shows a decline after $2^\circ.9$ radius at 57 km s$^{-1}$. At $4^\circ.3$ the $V_{rot}$ is $\sim70$ km s$^{-1}$ after which the fitted data seems to be sparse. We are unable to trace the flat part of the rotation curve with the coverage of ATCA/Parkes data. The de-projected rotation curve estimated by \citet{1998ApJ...503..674K} using H {\sc i} shows a sudden decline at a radius of $2^\circ.8$ and $V_{rot}$ of 63 km s$^{-1}$. \citet{1998ApJ...503..674K} attributes this decline to non-circular motions due to various reasons like interaction with the SMC and the MW, presence of bar, etc. \citet{2011ApJ...737...29O} estimated the rotation curve using 738 red super giants with a maximum value of $V_{rot}$ $87\pm5$ km s$^{-1}$ at a radius of $2^\circ.8$, (and goes flat beyond that), which is said to be consistent with the H {\sc i} rotation curve. \citet {2008AJ....135.1024P} presented a rotation curve with an amplitude of 120 km s$^{-1}$ at a radius of $4^\circ.6$ using Hubble space telescope data and \citet{2002AJ....124.2639V} estimated a value of 50 km s$^{-1}$ at a radius of $4^\circ.6$ using carbon stars. \citet{2014ApJ...781..121V} estimated the rotation curve of the LMC (their figure 6). Their estimate of the velocity at similar radius is found to be consistent with our estimates. 
\begin{figure}
   \resizebox{\hsize}{!}{\includegraphics{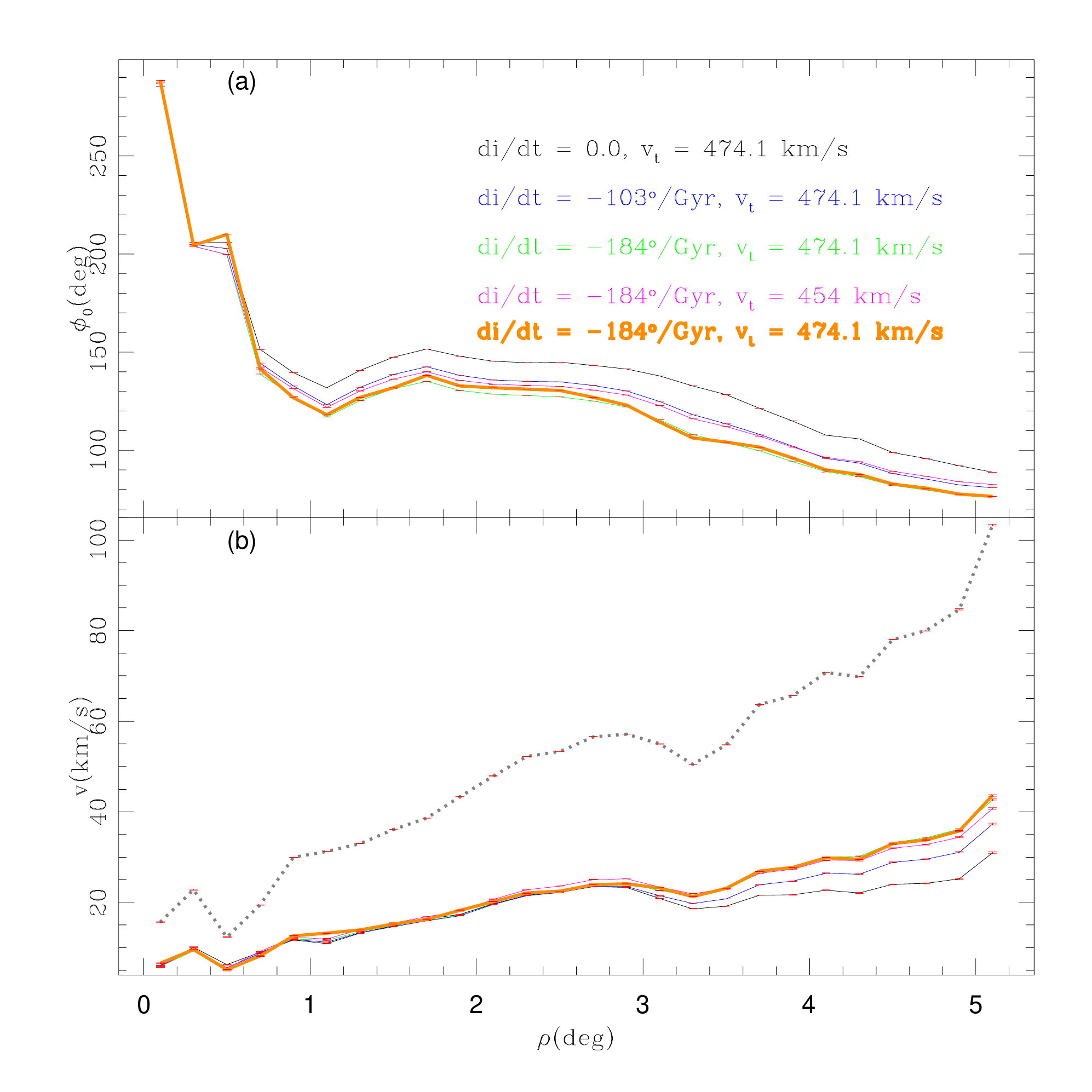}}
   \caption{Panel (a) shows the variation of PA as a function of radius. Panel (b) shows the radial variation of circular velocity. Five curves are shown for various $V_{t},di/dt$ values as mentioned in panel (a). The bold orange curve is from the first iteration after removing the outliers. The parameters corresponding to this curve is used for further analysis. The dotted curve in panel (b) is the de-projected rotation curve. The error bars are shown in red.}
    \end{figure}
\begin{table*}
      \caption[]{PA estimated for various radial extends using different $V_t,di/dt$ values. The last column is the estimated standard deviation in PA.}
         \label{Table:2}
	\centering
	\begin{tabular}{ c | c | c | c | c | c}
	\hline\hline
	 $rho(^\circ)$ & $Mean PA (^\circ)$ & $Mean PA (^\circ)$ & $Mean PA (^\circ)$ &  $Mean PA (^\circ)$ & $PA_{SD} (^\circ)$\\
	& $V_t = 474.1 km/s$ & $V_t = 474.1 km/s$ & $V_t = 474.1 km/s$ & $V_t = 454 km/s$ &\\ 
	& $(di/dt = 0^{\circ}.0/Gyr)$ & $(di/dt = -103^{\circ}/Gyr)$ & $(di/dt = -184^{\circ}/Gyr)$ & $(di/dt = -184^{\circ}/Gyr)$ \\
	\hline
	0.0 - 5.1 & 140 & 131 & 125 & 130 & 45 \\
	\hline
	0.5 - 5.1 &  132 & 121 & 115 & 120 & 28\\
	\hline
	0.5 - 4.0 &  143 & 133 & 126 & 131 & 23\\
	\hline
\end{tabular}
   \end{table*}
\subsection{Modelling the H {\sc i} disk}
\hspace{4ex}We used the parameters estimated, (before doing the de-projection but after the first iteration, the bold orange curve in figure 3) for the 29 annular rings for further analysis. Using these parameters we computed the velocity map of the modelled H {\sc i} disk of the LMC, shown in figure 4. We used di/dt = $-184^\circ/Gyr$, and $V_t$ = 474.1 km s$^{-1}$ to estimate figure 4. 
Our map estimated by the method of annular rings, beautifully reproduces almost all the features of the observed map (figure 1), except for the dark blue stream of points representing the high velocity gas moving towards us, located in the south-east (SE) direction, and the dark red points in the periphery of the western side of the disk. We discuss these features in detail in the next section.\\
The modelled H {\sc i} disk of the LMC, as shown in figure 3, suggests a complicated kinematics. The inner regions are found to have different PA, when compared to the outer regions. The kinematics in the inner 1.$^o$0 is quite different in terms of velocity as well as PA. This might suggest a strong internal perturbation, which could be due to the presence of the bar.
The kinematics of regions outside the bar is found to be quite smooth. Between the radius of 1.$^o$0 and 2.$^o$9, the disk is found to have similar PA and the velocity is found to increase with radius. After 2.$^o$9 radius, the disk shows effects of perturbation as seen by a sudden kink in the velocity and a gradual decrement in PA. The outer perturbation seen in the disk kinematics may be caused by the LMC-SMC-MW interactions.
  
The modelled H {\sc i} disk we estimated in our analysis is expected to closely represent the mean H {\sc i} velocity field of the LMC, obtained from the intensity weighted velocity map. This modelled disk is thus devoid of all small deviations in H {\sc i} velocity, but will have the signatures of large scale or global deviations. This disk is likely to show the disturbance/effect of the LMC-SMC-MW interaction either tidal or hydrodynamical.
The annular ring analysis suggested the presence of deviations from the mean disk and these might be due to the above disturbance. Thus, in the following sections, we present the outlier maps, estimated as regions, which show significant kinematical deviations from the modelled velocity distribution. 
\begin{figure}
   \resizebox{\hsize}{!}{\includegraphics{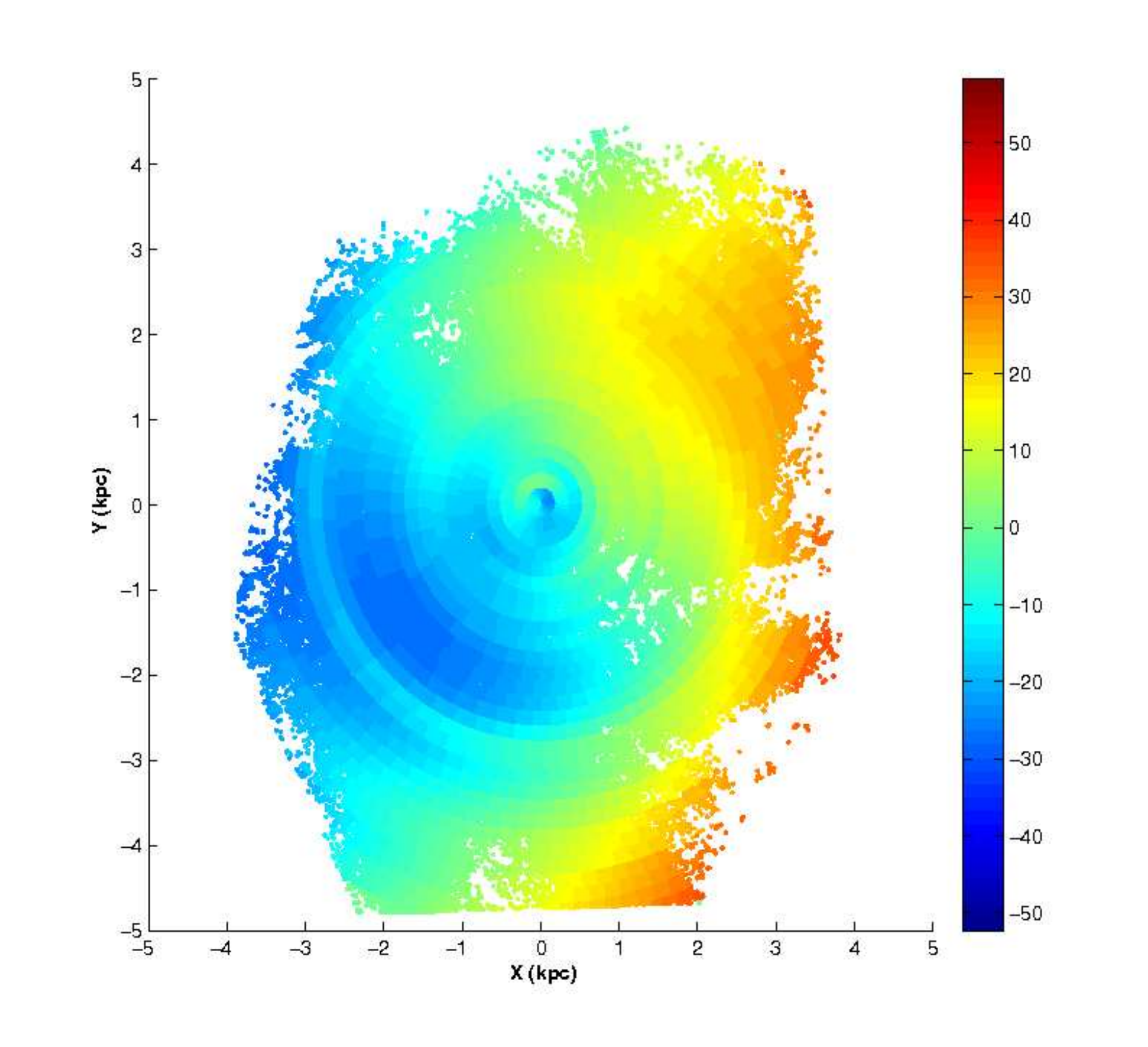}}
   \caption{Modelled H {\sc i} disk of the LMC. Colour coding is according to the variation in $V_{fit}$ in km s$^{-1}$.}
    \end{figure}
\section{Residuals \& outliers}
\hspace{4ex}After estimating the modelled H {\sc i} disk of the LMC, we find the distribution of kinematic outliers in this section.
The residual velocity is estimated as the difference between $V_{fit}$ \& $V_{mod}$.\\
\begin{equation}
V_{res} = V_{mod} - V_{fit},\\
\end{equation} 
where $V_{res}$ ranges from $\sim$ $-$50 to $+$50 km s$^{-1}$. Figure 5 shows the X-Y plots for nine various slices of $V_{res}$ with an interval of 10 km s$^{-1}$. There are only very few points ($<$ 1 \% of the total data) in panels a and b with V$_{res}$ $>$ $+$30 km s$^{-1}$ and in panel, i,  with V$_{res}$ $<$ $-$30 km s$^{-1}$. Panels c and d have an occupancy of 0.7 and 9.4 \% of data points, respectively.  Panel e has 39.8 \% and panel f has 41.7 \% as more than 80 \% of the data points have V$_{res}$ in the interval $-$10 to $+$10 km s$^{-1}$. Points in these panels cover almost the whole LMC disk except for a few locations. Further, panel g has 7.9 \% of data points and panel h has 0.6 \%. Approximately, less than 20 \% of the data points are likely to be kinematically deviated from the mean disk of the LMC. 
To distinguish the kinematic outliers,  we estimated the distribution of the velocity dispersion about the mean disk velocity. To estimate the velocity dispersion, we binned the velocity data, in sub-regions of area 17$^\prime$ x 17$^\prime$, in the projected X-Y plane. The number of velocity data points in the sub-regions varied from 1 to 2500 and the maximum Poisson error in this estimation is 50. A statistical cut-off of 3$\sigma$ is applied, so that the sub-regions with less than 150 data points are not considered for the analysis. For each selected sub-region, the mean and standard deviation of the first moment velocity is found. The statistical distribution of this estimated velocity dispersion (${\sigma}_V$) is shown in figure 6, which shows a maximum value of about 13 km s$^{-1}$ with a bin width of 2 km s$^{-1}$. About 94.2 $\%$ of the data points are within a ${\sigma}_V$ of 10 km s$^{-1}$. Therefore, for further analysis, the maximum value of ${\sigma}_V$ is taken to be 12 km s$^{-1}$, including the bin width of 2 km s$^{-1}$. The dispersion distribution is found to be narrow as expected for a kinematically cold system and has a single peak at 6 km s$^{-1}$.  \citet{1999AJ....118.2797K} estimated the velocity dispersion to be between 6.8 and 7.7 km s$^{-1}$ and adopted a mean value of 7.3 km s$^{-1}$ for the vertical velocity dispersion of the gas. Hence the two estimations of velocity dispersion match fairly well. In the ${\sigma}_V$ map (figure 7), we could identify locations with higher value of ${\sigma}_V$ around 12 km s$^{-1}$ or more near the south-east H {\sc i} over-density (SEHO) region, 30 Doradus and constellation III. Also, there are regions with large deviation near the centre and south-west (SW) part of the centre. \\
\begin{figure}
  \resizebox{\hsize}{!}{\includegraphics{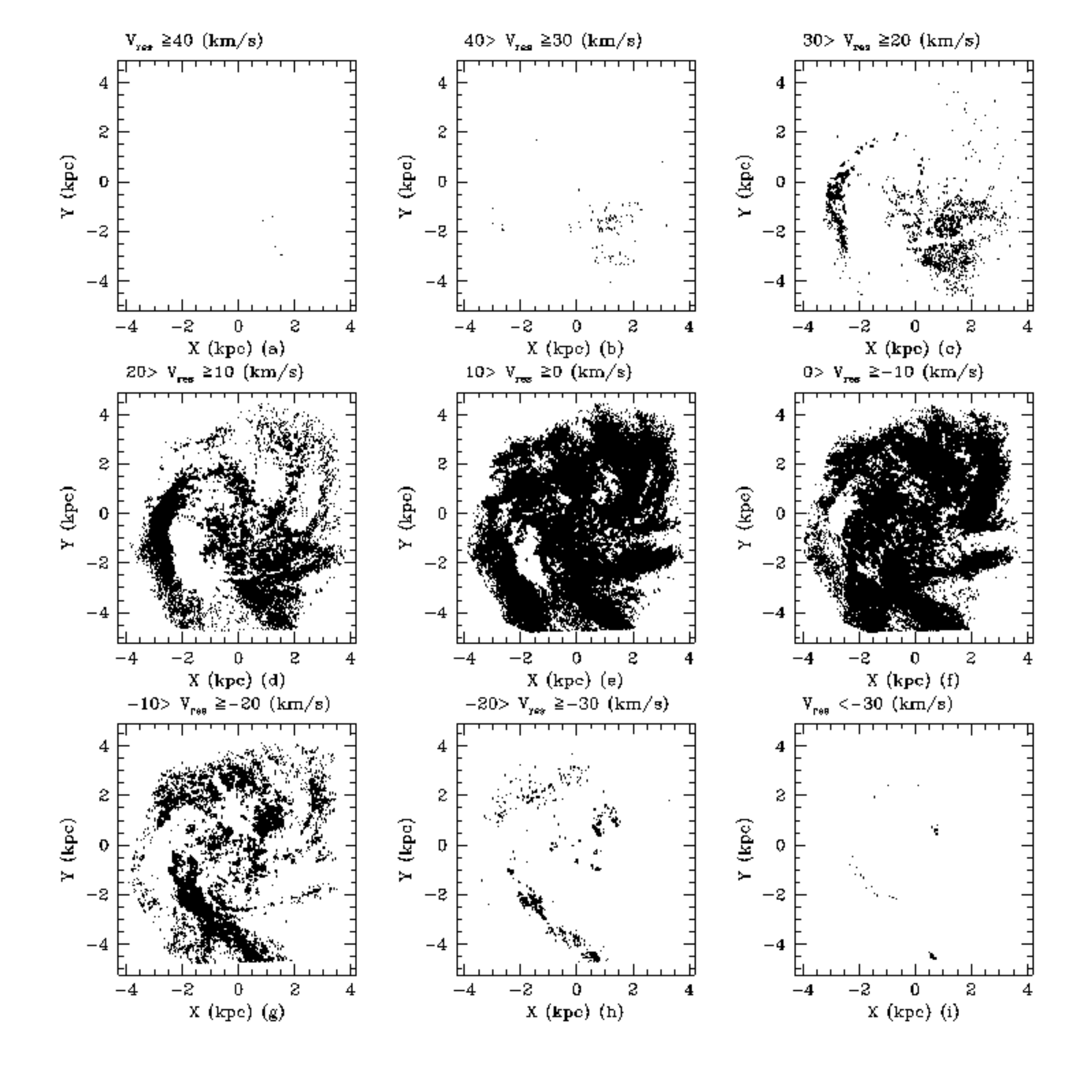}}
   \caption{Distribution of $V_{res}$ in the XY plane is shown in nine slices. The nine panels (a) to (i) show various slices in $V_{res}$, with the interval shown in the top of each panel.} 
    \end{figure}
\begin{figure}
   \resizebox{\hsize}{!}{\includegraphics{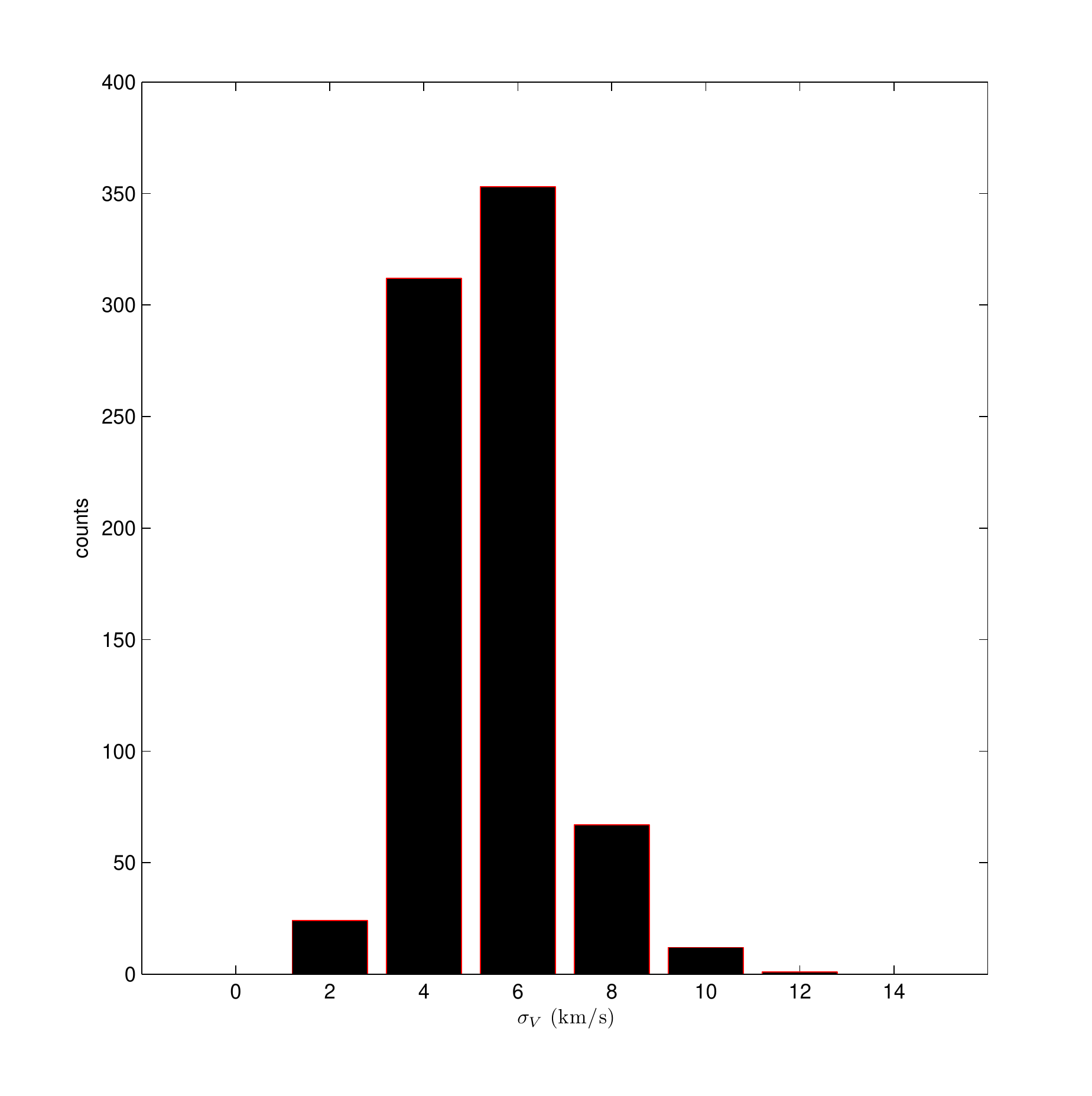}}
   \caption{Statistical distribution of the velocity dispersion ${\sigma}_V$.}
    \end{figure}

\begin{figure}
   \resizebox{\hsize}{!}{\includegraphics{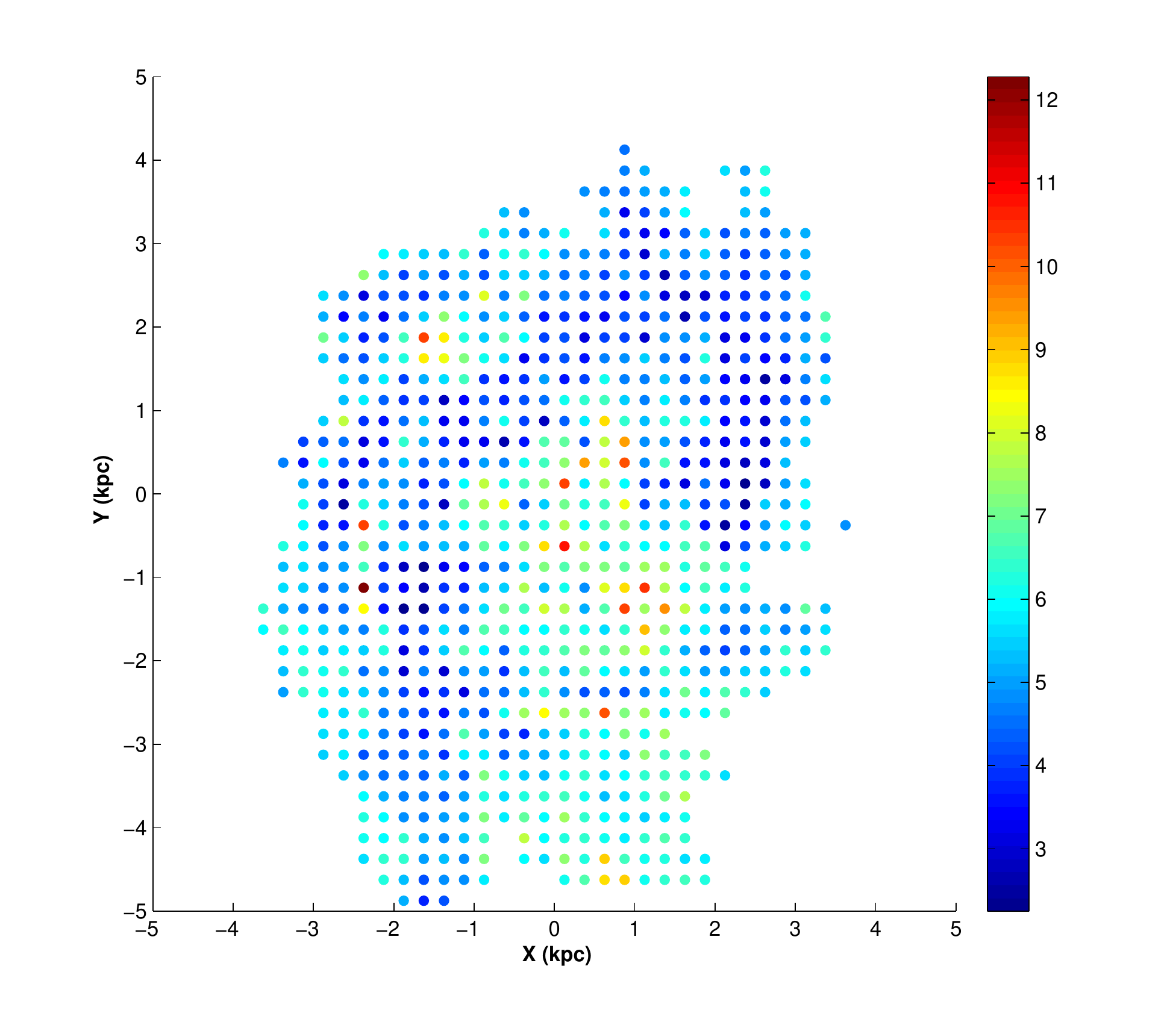}}
   \caption{Map of the velocity dispersion. Colour coding is according to the variation in ${\sigma}_V$ in km s$^{-1}$.}
    \end{figure}

Data points, which have $V_{res}$ $>$ 12 km s$^{-1}$ (maximum value of ${\sigma}_V$ + bin width), are defined as kinematic outliers. In the total set of H {\sc i} data points, 12.1$\%$ are identified as outliers using the above criterion. The fact that about 88\% of the data points are within the modelled disk suggests that the modelled disk is indeed close to the true disk of the LMC and the deviating points are a minority. These kinematic outliers can be classified widely into two categories, the type 1 outliers, where the observed velocity $V_{mod}$ bears opposite sign compared to the expected $V_{fit}$ and the type 2 outliers wherever the $V_{mod}$ and $V_{fit}$ have the same sign for velocity.  The type 2 outliers can be further classified into fast and slow components. In other words, if the $V_{mod}$ $>$ $V_{fit}$ by 12 km s$^{-1}$ or more, and if both $V_{mod}$ \& $V_{fit}$ bear same sign, then we define it to be a fast type 2 kinematic outlier. In the same manner, if $V_{mod}$ $<$ $V_{fit}$ by 12 km s$^{-1}$ or more and both have same sense of rotation, then we identify it as a slow  type 2 outlier component. If the magnitude of the $V_{res}$ $>$ 12 km s$^{-1}$ and if $V_{mod}$ and $V_{fit}$ bear opposite signs then the location is identified to be type 1. 
The present model is valid only up to a radius of 4$^o$, hence the differentiation of an outlier to be type 1 or type 2 is possible only within this radius. The locations of these type 2 and type 1 data points are identified and addressed in detail in the following sections.

\subsection{type 1 kinematic outliers} 
\hspace{4ex}We found around 2.7$\%$ of the total data sample to be type 1 outlier, the map of which is shown in figure 8. In the eastern side of the disk, the type 1 gas forms an arm-like feature (the pale green stream of points extended vertically from north to south, marked as E in figure 8), very similar in location to Arm E \citep{2003MNRAS.339...87S}. This feature is also visible in figure 5 in panel c with $V_{res}$ ranging from $+$30 to $+$20 km s$^{-1}$.
If we examine figure 8, the same component is found to continue to the north and north-west (NW). The data points here have a velocity around $-$10km s$^{-1}$, whereas the modelled disk rotation velocity is about 25 km s$^{-1}$. It is tempting to extend the Arm E up to the NW. \citet{2008ApJ...679..432N} observed Arm E to emanate from many outflows across the SEHO region. \citet{2011ApJ...737...29O} also traced a counter-rotating stellar component to be similar in location to Arm E and concluded that Arm E may not be coming out of the LMC, but probably falling in. We suggest that Arm E is part of the type 1 component in the LMC, which has a velocity very close to the systemic velocity. 
Another feature of interest is the component near the southern end of Arm E at X=$+$0.5, Y=$-$4.5 having a velocity around 10 km s$^{-1}$ towards us (the light blue points).  This feature is at a radius of about 5$^\circ$, and we cannot classify it reliably. 


 There are a few more specific locations in figure 8 where we found regions that are spatially concentrated, but cover a large range in velocity. One such location is around X=$+$0.8, Y=$+$0.6, where we see a concentration of dark blue points surrounded by light blue points. If we trace the same location in the ${\sigma}_V$ map (figure 7), it shows a relatively high standard deviation in velocity. A similar location can be found near constellation III (the small black dot at X=$-$1.5 and Y=$+$2 in figure 8). The location of SGS as well as giant-shells (GS) (from \citealt{1999AJ....118.2797K}) are shown in figure 8. Only a few SGS and some GS are located along the blue/green points distributed between the above two locations. In general, the locations of SGS as well as GS are not coincident with the type 1 regions.
\begin{figure}
  \resizebox{\hsize}{!}{\includegraphics{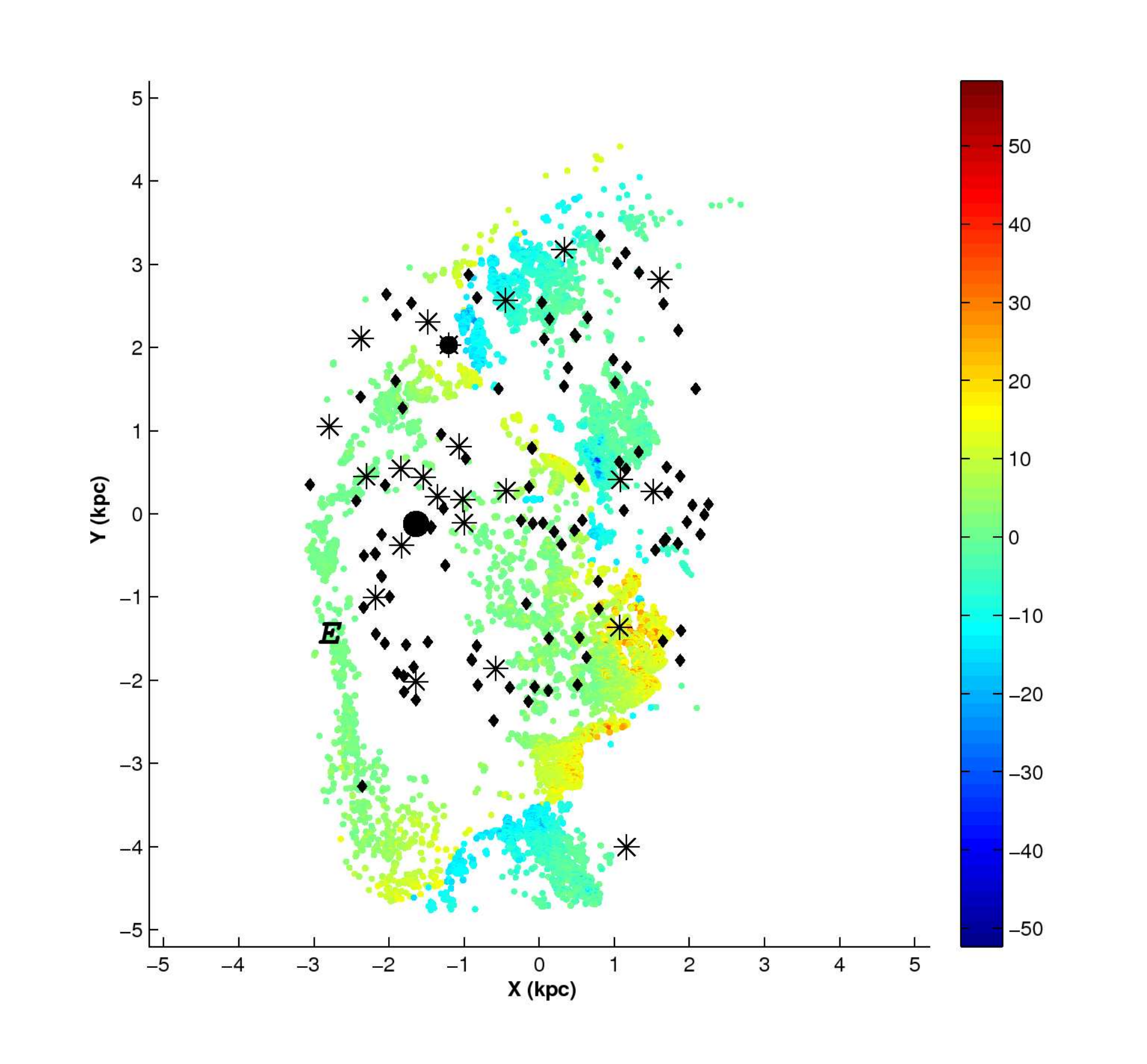}}
   \caption{Map of the kinematic outliers identified as type 1. Colour coding is according to the variation in $V_{mod}$ in km s$^{-1}$. The big black dot represents the location of 30 Doradus and the small dot is constellation III. The locations of H {\sc i} SGS are shown as stars and the locations of H {\sc i} GS are shown as diamonds (from \citealt{1999AJ....118.2797K}). The possible location of Arm E is marked. The corresponding $V_{res}$ ranges from $\sim$ $-$48 to $+$45 km s$^{-1}$.}
    \end{figure}
\subsection{Slow type 2 kinematic outliers} 
\hspace{4ex}The kinematic outliers, which have the same sign of velocity, (type 2) can be classified into two groups. One with apparent faster rotation, which is called fast type 2, and the other with apparent slower rotation, which is called slow type 2. Figure 9 shows the map of slow type 2 outlier component. The slow component, which constitutes 3.9$\%$ of the H {\sc i} data, has velocities in the range $-$14 to $+$16 km s$^{-1}$. The narrow statistical distribution has a single peak at around $-$6 km s$^{-1}$, which constitutes around 0.8$\%$ of the total data points. The slow type 2 component is accumulated mostly in the SE of the LMC disk. The light blue stream of points located vertically NS, in the SE part of the disk, closely resembles the location of Arm E (marked as E, in figure 9). This feature is also visible in figure 5 in panel d with $V_{res}$ ranging from $+$20 to $+$10 km s$^{-1}$.
As its location is coincident with the location of type 1 gas (Arm E), 
 it is possible that slow type 2 and type 1 are connected. 
In figure 9, we also notice the presence of a slow rotating component in the central as well as in the NW regions. As the southern most point is outside the radius of 4$^o$, we cannot reliably classify them.
\begin{figure}
  \resizebox{\hsize}{!}{\includegraphics{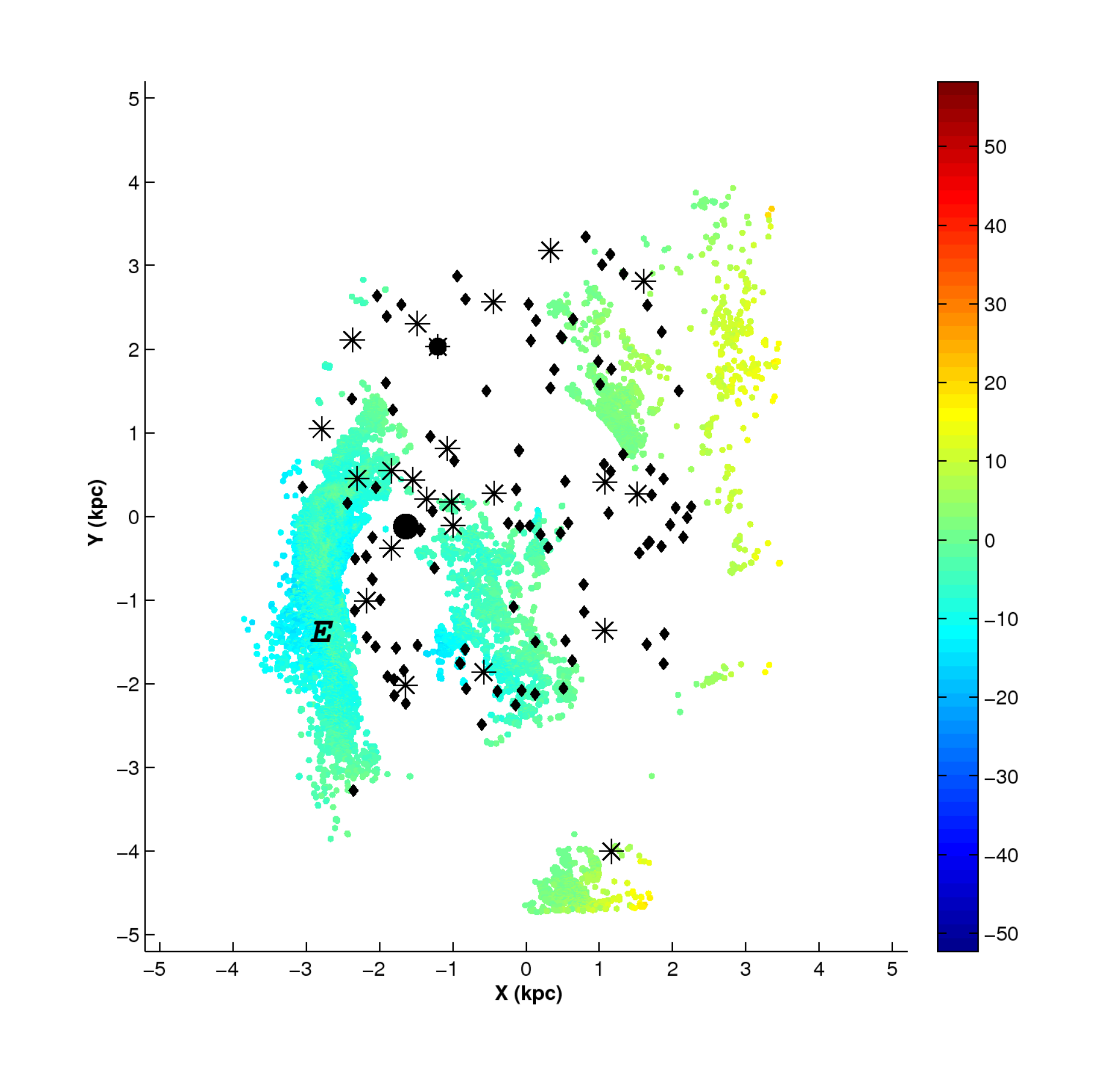}}
   \caption{Map of the slow type 2 kinematic outliers. Colour coding is according to the variation in $V_{mod}$ in km s$^{-1}$. The big black dot represents the location of 30 Doradus and the small dot is constellation III. The locations of H {\sc i} SGS (from \citealt{1999AJ....118.2797K}) are shown as stars. The possible location of Arm E is marked.  The corresponding $V_{res}$ ranges from $\sim$ $-$23 to $+$26 km s$^{-1}$.}
    \end{figure}

\subsection{Fast type 2 kinematic outliers}
\hspace{4ex}The velocity map of the fast type 2 outlier component, which is nearly 5.5$\%$ of the total H {\sc i} data, has $V_{mod}$ ranging from $-$50 to $+$60 km s$^{-1}$ as shown in figure 10. The statistical distribution is not continuous and has prominent peaks at around $-$32 to $-$30 km s$^{-1}$, at $-$40 km s$^{-1}$, which constitute around 0.3$\%$ and 0.4$\%$ of the total data points, respectively, and smaller peaks at $\pm$20 km s$^{-1}$, which constitutes around 0.2$\%$ each. We found a stream of fast type 2 H {\sc i} gas (the dark blue stream of points in the SE, marked as B in figure 10) coincident with Arm B \citep{2003MNRAS.339...87S}. This feature is seen in figure 5 in panels, g and h. The $V_{res}$ in this feature ranges approximately from $-$10 to $-$30 km s$^{-1}$. One can also notice in panel e, a prominent gap in the central region, caused by these data points. In figure 10, this feature is found to disappear near the 30 Doradus region (close to X=$-$1.5 and Y=0), shown as a big black dot. At this location, one can also see one section of the structure like digit 8, as seen in figure 1 of \citet{2003MNRAS.339...87S}. Also, this feature splits up near the central LMC. As shown in \citet{2003MNRAS.339...87S}, Arm B is pointed towards the MB/SMC. If the above feature is similar to Arm B and connected to the MB, then this might suggest a connection of gas from the MB/SMC to the central LMC near 30 Dor. \citet{2013A&A...550A.108V} suggested a possibility of 30 Dor region accreting cool gas. \citet{2008ApJ...679..432N} found a gas stream (figure 12, \citealt{2008ApJ...679..432N}) which they called the LMC filament, which is associated with Arm B and emanates from the SEHO region. 
 
 We also notice another less prominent arm-like structure resembling Arm B,  but found in the eastern part of Arm B. This stream can be traced from X=$-$2 and Y=$-$4.5 up to X=$-$3, Y=$+$1, close to the location of the SGS in the NE. This is a newly identified structure and we call it the 'Outer Arm'. (This is marked in figure 10). This feature is clearly seen in figure 5, mostly in panel g, and minimally in h. The $V_{res}$ in this feature also ranges approximately from $-$10 to $-$30 km s$^{-1}$. From figure 10, this stream seems to be located to the east of Arm B and located in the outer part of the LMC disk. In the NE, the stream gets spread and we see a large area occupied by this fast type 2 component (a pool of blue points). This is also the region that is expected to be compressed due to the motion of the LMC in the halo of the MW (at a PA $\theta_t$ $\sim$ 80$^\circ$, calculated as the average of $\theta_t$ values from table 2). 
 We also notice the presence of a few SGS coincident with this location. On the other hand, a stream can be traced to continue beyond this pool of blue points, towards the north. 
This stream changes velocity as it moves towards the west. This change over from the blue points to the red points could suggest a change in the direction of motion. That is, the gas which was moving towards us, now starts moving away from us, similar to a change over across the minor axis of a rotating disk. 
Near the SGS in the NW side of the disk (at X=$+$2, Y=$+$3), the stream splits into two, one moves to the inner part of the disk and the other stream continues in the outer radius. The inner stream (marked as W in figure 10) is found to be similar to Arm W as identified by \citet{2003MNRAS.339...87S}. 
This stream then can be traced up to the SW corner, where it joins the large area occupied by the high velocity gas. 
Beyond this point, considerable spread in the velocity as well as the location of the stream can be noticed, suggestive of the gas getting dispersed. The locations of many of the SGS fall in the path of the identified fast type 2 kinematic outliers, suggesting that this gas, moving faster than the disk, could be interacting with the gas in the disk resulting in vigorous star formation and the creation of such shells. In figure 5, arm W is seen in panels d and e, in the $V_{res}$ range $+$20 to 0 km s$^{-1}$. 
In figure 10, we detect high velocity (red) points in the SW/south of the LMC disk and these are suggestive of gas moving away from us, much faster than the disk. 
The location of this high velocity gas in the disk is in a region diagonally opposite to the direction of movement of the LMC. 
In figure 10, we also identify a faint feature, similar to the Arm S \citep{2003MNRAS.339...87S} at X=0, Y=$-$2.5, (marked as S in figure 10) located to the south of the high velocity gas (pool of red points) near the SW side.
The shape and curvature of Arm S is very different and it does not appear to be connected to any other features. This feature is clearly seen in figure 5, panel g with $V_{res}$ $-$10 to $-$20 km s$^{-1}$.  
\begin{figure}
  \resizebox{\hsize}{!}{\includegraphics{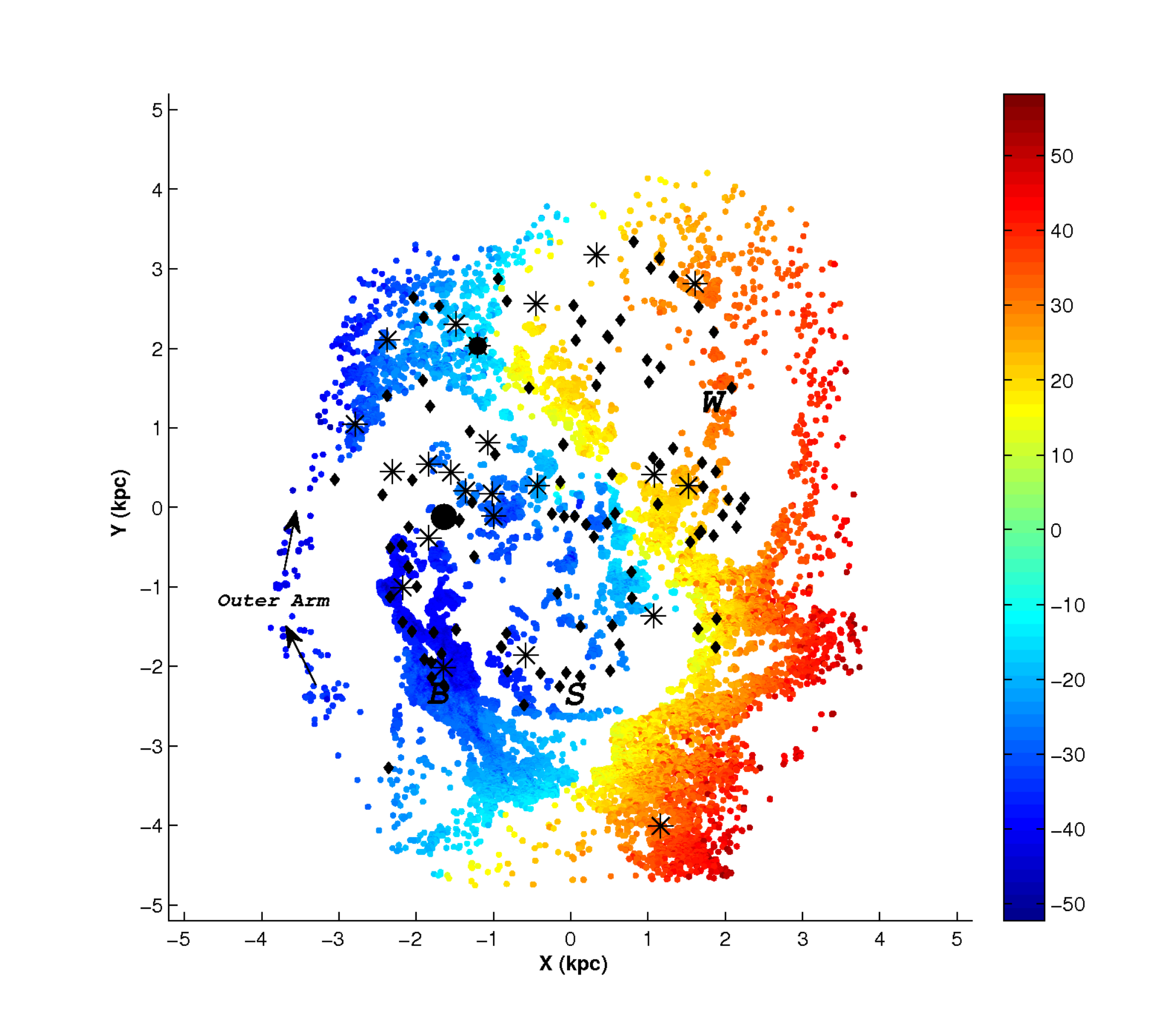}}
   \caption{Map of the fast type 2 kinematic outliers. Colour coding is according to the variation in $V_{mod}$ in km s$^{-1}$. The big black dot represents the location of 30 Doradus and the small dot is constellation III. The locations of H {\sc i} SGS (from \citealt{1999AJ....118.2797K}) are shown as stars. The Outer Arm, is marked along with arms B, W \& S. The corresponding $V_{res}$ ranges from $\sim$ $-$36 to $+$42 km s$^{-1}$.}
    \end{figure}
 
 We thus identified and located three types of H {\sc i} kinematic outliers in the H {\sc i} gas present in the LMC disk. The type 1 and the slow type 2 outliers are mostly located in the SE region, which are identified as part of Arm E. The Arm E is known to connect to the LA. Arm B, Arm W and Arm S are detected as fast type 2 outliers and we also identify an eastern arm called Outer Arm. Also, the Arm B and the Outer Arm point to MB/SMC. We need to understand the possible origin of these outliers. 
The ATCA/Parkes data does not cover the outer regions of the LMC and hence we are unable to trace various features to the outer disk of the LMC. In order to study their extensions, we have analysed the GASS data set, which covers a larger area. 
\section{GASS data analysis}
\hspace{4ex}We estimated the intensity weighted mean velocity map for the GASS data set. For this purpose, the intensity weighted average of the $V_{los}$ is calculated for each pixel. These first moment maps provide the best estimate of the mean velocity field of the LMC extractable from the data set. We corrected the line-of-sight velocity field for the systemic, transverse, precession, and nutation motions of the LMC to obtain the distribution of $V_{mod}$.
The annular ring analysis is not performed with the GASS data set because of its lesser spatial resolution compared to ATCA/Parkes data. The kinematic parameters of the mean disk estimated using ATCA/Parkes data is used to analyse the GASS data set as well. For the GASS data, the parameters $V$, $\delta V_{sys}$ and $\phi_0$ estimated using ATCA/Parkes data set is used and the expected velocity distribution is estimated. The $V_{res}$ values are estimated as the difference between $V_{mod}$ \& $V_{fit}$ using equation 3. 
In order to identify the kinematically distinct features, points that have $V_{res}$ $>$  12 km s$^{-1}$ are defined as outliers for the GASS data. The outliers are classified in the same way as done previously. As the estimated mean disk using ATCA/Parkes data is spatially restricted within a radius of $\sim5^\circ$ within the LMC, we assumed the disk to be flat beyond this radius. So it is ambiguous to classify the outliers to be type 2/type 1 beyond this radius. 
 \\
\begin{figure}
   \resizebox{\hsize}{!}{\includegraphics{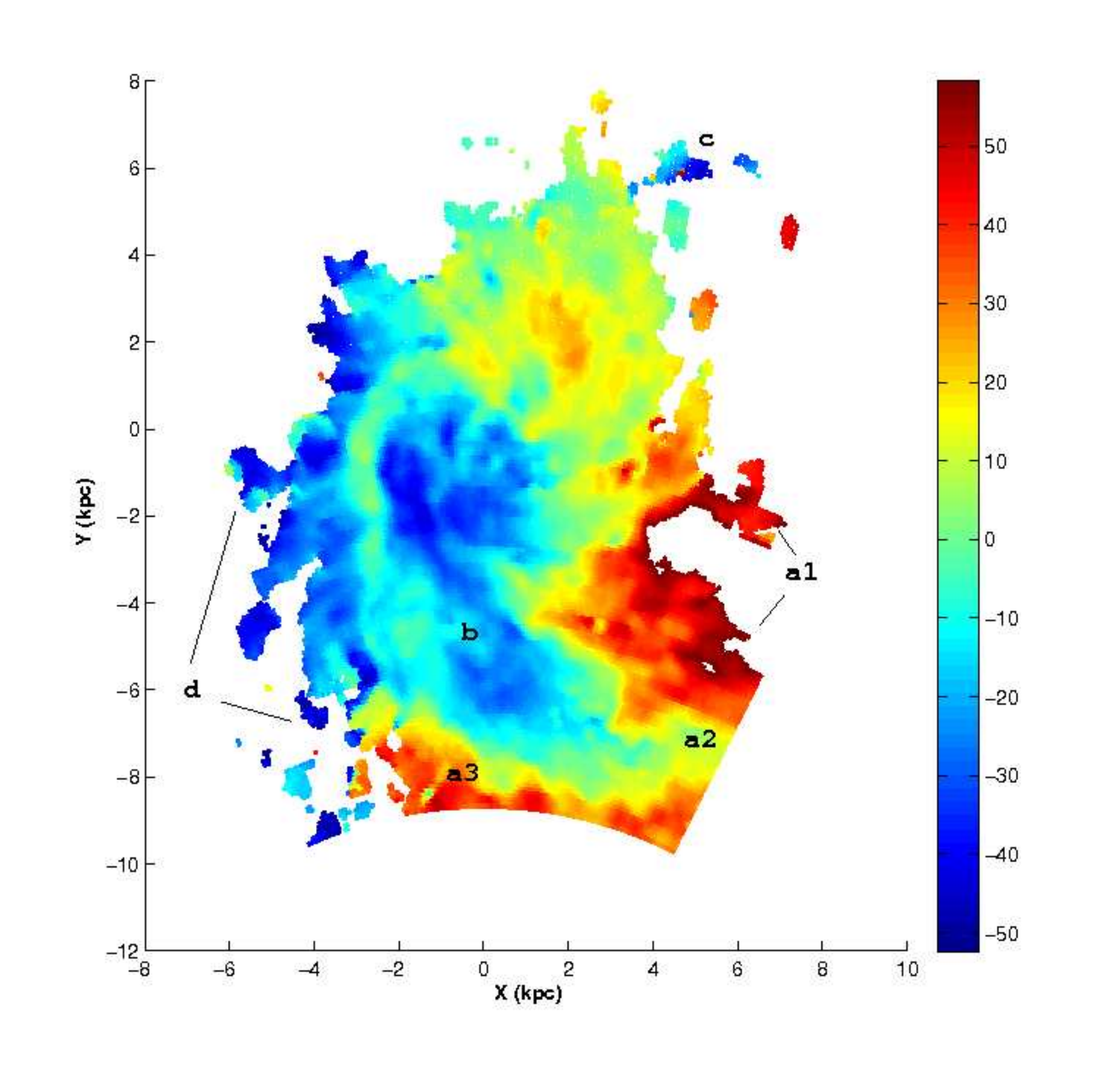}}
   \caption{Intensity weighted mean velocity map using GASS data. Colour coding is according to the variation in $V_{mod}$ in km s$^{-1}$. The alphabets marked are : the high velocity red points located in the west and SW of the LMC disk, (a1); the high velocity red points located in the south and SW of the MB, (a2) and (a3) respectively; the extension of Arm B feature to the MB/SMC, (b); the plume of blue points near the NW edge, (c); and the blue points in the south and SE of the LMC disk, (d).} 
    \end{figure}
\subsection{The GASS H {\sc i} disk and the distribution of outliers.}
\hspace{4ex}Figure 11 shows the intensity weighted mean velocity map of the LMC using GASS data, after applying the corrections mentioned earlier. This map using GASS data is generated for the first time. This is a spectacular H {\sc i} map of the LMC, which covers a large area, thus including the MB. The features seen in the data are not affected by any assumptions on the LMC disk, and the line-of-sight velocity field is corrected for systemic, transverse, precession, and nutation motions of the LMC. Thus the details identified in the outer regions of the LMC disk as well the MB are true features.
In the inner regions up to a radius of 4-5$^o$, the features seen are very similar to figure 1. Outside this radius, we can see that the GASS data goes quite farther, particularly in the 
south and SW direction. The most prominent features are the MB, which is connected to the LMC disk and its large extent. This figure brings out the details of the connection between the LMC disk and the MB. We also notice the large width (angular extent) of the MB apart from its radial extent. These extensions can help us to connect various features identified using the ATCA/Parkes data. One of the main
features is the red points towards the west and SW end of the LMC disk (a1 in figure 11). These high velocity red points are suggestive of gas moving away from the LMC. Another main feature is the extension of Arm B to the MB (b in figure 11), which in turn is connected to the SMC. There are locations similar to a1 in the south and SW of the MB (a2 and a3 in figure 11) where we can observe high velocity red points. If we proceed radially inward through the SW feature (a2 in figure 11), we could notice that the velocity changes from red to blue, suggesting a change from high positive velocity to negative velocity. As this feature is part of the MB, this suggests that the gas in the inner part of the MB has velocity opposite to the outer part. We also notice a plume of blue points near the NW edge (c in figure 11) as well as in the east (d in figure 11).

We now identify the kinematic outliers in the GASS data as shown in figure 12. 
We are able to trace all the previously identified outliers using ATCA/Parkes, in the inner region, though the details of the features are missing due to poorer resolution. The extent of this figure is ideal to connect the inner features to the outer regions, which was absent in the ATCA/Parkes data. Most importantly, the Arm B is found to be connected to the MB, whereas we could only suggest that it was pointed towards MB using the ATCA/Parkes data. The gas pool seen in the MB is found to be connected to the LMC through Arm B (blue points analogous to b in figure 11). Along Arm B, we observe a negative velocity with a gradient that tends to be more negative towards the 30 Dor region. The velocity, and the sign of velocity of this feature suggests that it is moving faster than the mean disk as it is part of the fast type 2 outlier. According to the inclination of the LMC disk with its NE part towards us, this velocity points to, gas flowing towards the LMC disk and not away from it. Hence the Arm B which connects the MB to the LMC disk (near 30 Dor region) could be an infall feature, but unlikely to be an outflow feature. This possibility is modelled and discussed in detail in the next sections. On the other hand, the region of the MB closer to the SMC has high positive velocity, which in turn suggests that the gas may be leaving the system from these locations. 

 The large velocity gradient observed along the MB suggests that the MB is being sheared. The gas closer to the LMC has a negative velocity and hence could fall into the LMC, whereas it can leave the galaxy from the other end as indicated by the positive velocity. We also notice high velocity gas, which could leave the LMC from the western side of the disk (red points analogous to a1 in figure 11). As the LMC moves in a PA $\sim$ 80$^\circ$ in the MW halo, the ram pressure effects can cause stripping of gas in a diagonally opposite direction. Thus, in the west and SW regions, the high velocity gas is expected to leave the disk. The Arm E is clearly seen in green and yellow points, which are analogous to the type 1 and slow type 2 components identified using ATCA/Parkes data. 
The blue points located to the east of Arm E are part of the newly found Outer Arm, which 
is connected to the MB/SMC. Also, the blue points can be noticed at large radii in the east and SE regions. This stream is found to terminate at a similar location in the NE, as in figure 10. In the western side, we notice the arm W and a feature suggestive of arm S. We also notice a blue plume of points in the NW region (location analogous to c in figure 11), which is part of the type 1 feature identified using ATCA/Parkes data. 
\begin{figure}
   \resizebox{\hsize}{!}{\includegraphics{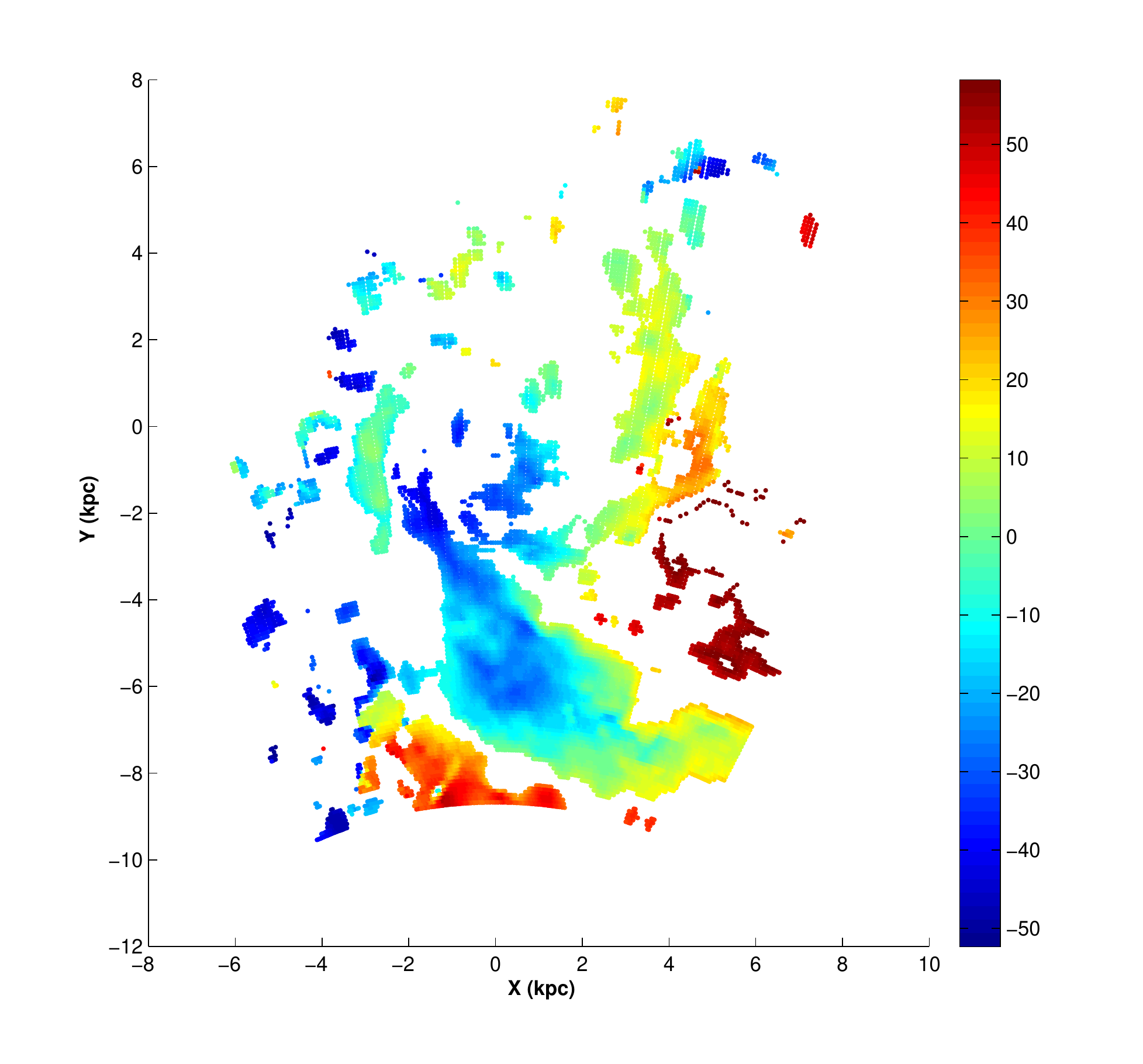}}
   \caption{Outlier component map identified using GASS data. 
Colour coding is according to the variation in $V_{mod}$ in km s$^{-1}$. The corresponding $V_{res}$ ranges from $\sim$ $-$67 to $+$67 km s$^{-1}$.}
    \end{figure}

 In summary, the GASS data analysis has revealed the following features.  There is an evident connection of the LMC to the MB through Arm B and possibly through the Outer Arm. 
We identify a possible connection between the Arm E and a gas pool in the NW. High velocity gas is identified in the western side of the disk and the south and SW of the MB, which is not seen clearly in the ATCA/Parkes maps. 

To explain the distribution of type 1 and type 2 outliers identified using ATCA/Parkes data, and the features related to these outliers identified using ATCA/Parkes as well as GASS data sets, we have come up with two possible scenarios, which are discussed in the following sections.

\section{Origin of outliers}
\hspace{4ex}In this section, we have put forward two scenarios to explain the possible origin and current distribution of all the three kinematic outliers. The scenarios could also successfully explain some of the observed features in the LMC disk.
 
\subsection{Scenario I : outlier gas in the plane of the LMC disk}   
\hspace{4ex}In scenario I, we consider that all gas, including the outlier gas is present in the same plane of the LMC disk. This model is based on the fact that the LMC disk and the line of interaction of the L\&SMC centres are coplanar (subtends an angle $\sim$ 5$^\circ$). We also assume the disk of the LMC and the MB are in the same plane. Hence according to the scenario I the outliers observed are true deviations in velocity. Figure 13 shows the LMC disk, where the sense of rotation is marked. With an inclination of around $25^\circ$ and with the NE part of the disk closer to us, disk of the LMC is found to have a clock-wise rotation. The green arrow shows the direction of the transverse motion of the LMC towards NE. As we discussed in the previous section, as the LMC moves in the MW halo at a PA $\sim$ 80$^\circ$, the gas is expected to strip away due to ram pressure effects, in the diagonally opposite side of the LMC disk. The expected direction of the ram pressure stripping is shown as a green arrow towards SW.
    
The analysis using GASS data suggests that gas in the LMC disk is presently connected to the MB. We identified two such streams Arm B and Outer Arm, which are fast type 2 outliers, which connect the LMC to the MB, only in the eastern side of the LMC. The outlier gas in Arm B can behave in three ways. The gas could be moving into the LMC as an infall feature, this gas would show up with velocity similar to the disk or faster (as in fast type 2). The gas could be stationary, which can be identified as gas with velocity close to the systemic velocity (as in slow type 2). If the gas in the Arm B represents an outflow, then the observed velocity will have the opposite sign (as in type 1). The high negative velocity and the velocity gradient detected for the gas in Arm B suggest that the gas may be moving into the disk along the Arm B as well as the Outer Arm. Thus, the gas velocity in Arm B and the Outer Arm (which are identified as fast  type 2) suggest that these are possible infall features from the MB to the LMC and not outflow or stationary features. Thus, for a case of the gas being located in the plane of the LMC, the schematic of a possible gas flow into the LMC from the MB is shown in figure 13 as black arrows. The arrows indicate the direction in which the gas would fall from the MB. The velocity of the infalling gas in the eastern side has the same sense of rotation as the LMC disk (the bottom arrow). We know that the SW side of the LMC is closest and well connected to the MB and hence the gas could also enter the LMC through the SW/western side of the LMC disk. On the other hand, if the gas falls in the western side, it will have the velocity directed opposite to the direction of the LMC disk rotation (the top arrow).  As the infalling gas sees only the gravitational potential of the LMC, and not the sense of rotation of the LMC disk, the gas is likely to move in either clock-wise or counter clock-wise direction at the point of infall. 
If the gas has low velocity, it is likely to move close to the centre directly, without spiraling in (the middle arrow). Based on this model I, we discuss whether we can produce all the kinematic components and explain their present distribution. 

As the velocity gradient suggests that there is a possible gas infall from the MB, we try to explain the identified features based on this infall. If the infalling gas reaches the western disk it moves in the counter clock-wise sense towards the north. While moving towards the north, it can get slowed down because of interaction with the disk gas moving in the opposite direction. Gas that is slowed down will move towards the centre in a path with lesser radius, which should be similar to what we see in figure 9, and we see a very similar distribution in figure 8. Gas that retains higher velocity would move towards the north. We are not able to detect this kind of stream in figure 8. On the other hand, we do see some blue dots near the two western SGS (at about X=$+$1 and Y=0), which are probably gas falling in like the top black arrow shown in figure 13. These points are also seen in the channel maps of \citep{2003MNRAS.339...87S} for velocities around 215 km s$^{-1}$. We also notice a large chunk of blue points in the NW, near (X=6, Y=6), in the GASS data (marked as c in figure 11), which suggests a connection to Arm E.
The large H {\sc i} density close to these two SGSs and the far NW point could be produced by the infalling gas in the velocity range $-$20 to $-$30 km s$^{-1}$. The H {\sc i} distribution in the eastern part (Arm E) identified as type 1 outlier can be produced by the faster component of this infalling gas. The gas as it moves counter clock-wise would get dragged/slowed down further by the gas moving in the clock-wise direction. This slowed down gas will eventually have velocity similar to the systemic velocity. This is probably what we see along Arm E, where the type 1 gas has velocity very close to zero, or systemic velocity. Once the velocity of the gas approaches zero velocity, then it can get dragged by the disk, which can produce the slow type 2 component. This explains the presence of gas in the eastern side as shown in figure 8 as well as 9. 
Extension of this feature to LA is dependent on the velocity of the gas accreted from the MB/SMC and its eventual slow down. Thus, gas with sufficiently high velocity located in the Arm E would leave the LMC, whereas the low velocity gas would fall back into the LMC.

 \citet{2005ApJ...625L..47S} first suggested the presence of counter-rotating stars and gas in the LMC. Recently, \citet{2011ApJ...737...29O} identified that this counter-rotating stellar population is accreted from the SMC. Thus, along with the gas, the accretion event could be bringing stars, from the MB as well.  
Thus, in our model I, gas with velocity ranging from systemic velocity up to $-$50km s$^{-1}$ with respect to the LMC disk falling into the west/SW disk of the LMC from the MB forms the counter-rotating component and can explain not only the type 1 as well as the slow type 2 kinematic outliers, but also the production of the LA. 
 
According to \citet{2009MNRAS.399.2004M}, the current orientation of the LMC disk with respect to its orbit is nearly edge on, whereas its orientation according to \citet{2007ApJ...668..949B}, is $-$30$^\circ$. Also the motion of the LMC in the Galactic halo is analogous to a wedge moving supersonically with the vertex facing upstream. As soon as the satellite moves in the surrounding medium, the external gas density develops a peak centred at a PA of the direction of its movement (In the case of the LMC disk this PA, $\theta_t$ is $\sim$ 80$^\circ$). The disk particles localised in regions of maximum ram pressure get compressed and move onto inner orbits while their circular velocity consequently increases. This may be the reason for the origin of high velocity gas, which moves along with the disk. The NE region is expected to be compressed due to LMC's motion. The high velocity streams, which we identify in the west, including Arm W,
might have been caused by the above mentioned effect. Also, the gas falling into the LMC will experience pressure due to the motion of the LMC in the MW halo. The gas coming in from the MB, might get pushed in, creating the Outer Arm. In the GASS data, we see diffuse gas towards the east of the Outer Arm. This gas will try to go around the LMC, like the Outer Arm, but may reach the north side of the disk due to the LMC motion.  Thus, gas with systemic velocity and up to about $-$50 km s$^{-1}$ with respect to the disk accreted in the eastern side gives rise to the fast co-rotating kinematic outlier, comprising of Arm B, the Outer Arm and the diffuse gas around the LMC disk.

 The red hexagons in figure 13 show the locations of high positive velocity points marked as a1, a2 and a3 in figure 11. According to the orientation and sense of rotation of the LMC disk, these locations are possibly where the gas could leave the system. As these coincide with the location of expected ram pressure stripping, it is likely that the high velocity gas at these locations leave the LMC due to ram pressure effects.
 
 Coplanarity or near-coplanarity of all gas, including the outliers, is the main assumption of this model, such that the gas connected to the LMC from the SMC, through the MB, is in the plane of the LMC. The assumption is based on the fact that the inclined disk of the LMC and the line connecting the centres of the L\&SMC makes a small angle $\sim$ $5^\circ$ (this is visualised in figure 14 in a ZY plane). Such an in-plane assumption along with the observed gradient in velocity and the sense of rotation of the LMC, suggest Arm B to be a possible infall feature and Arm E to be a possible outflow feature. 
\begin{figure}
  \resizebox{\hsize}{!}{\includegraphics{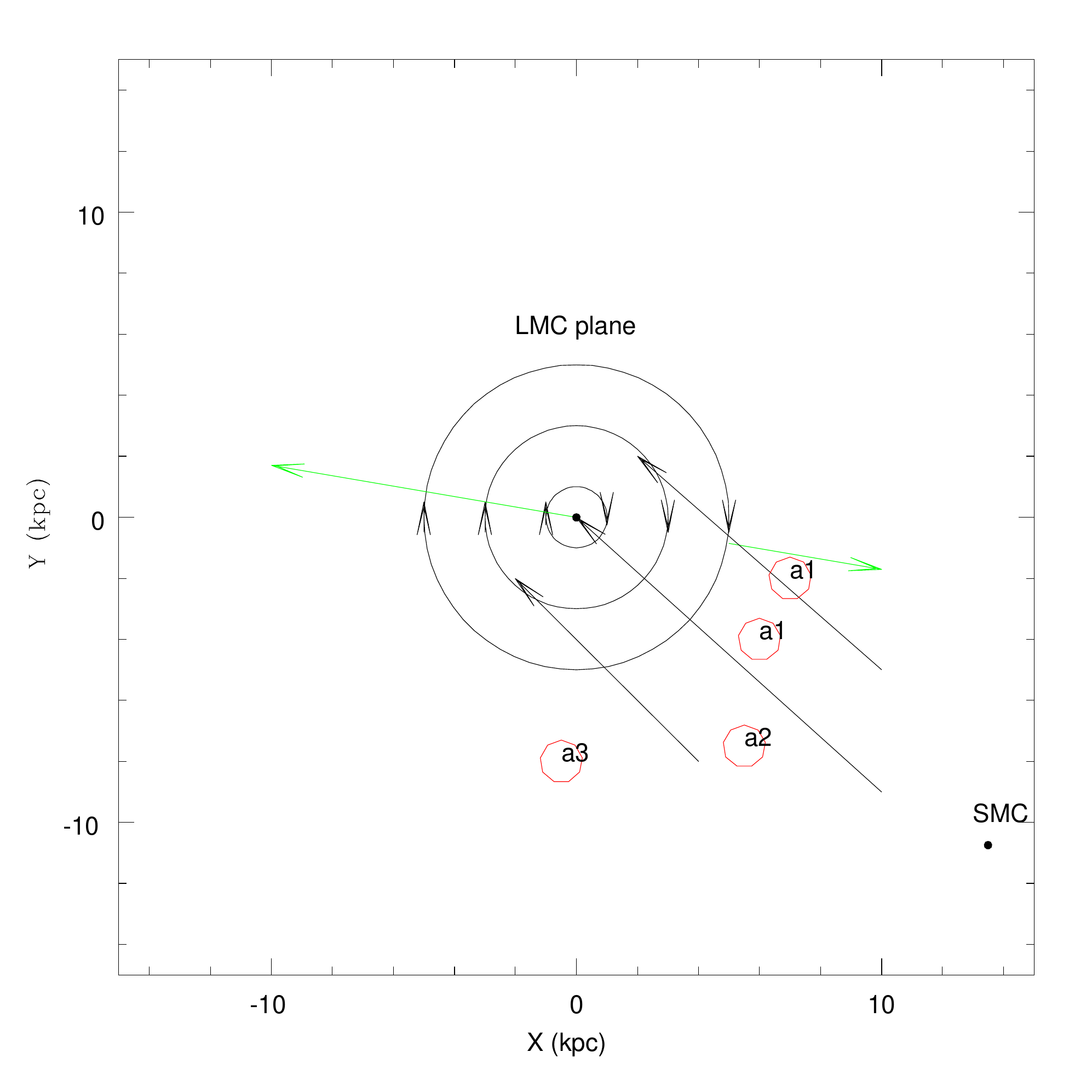}}
   \caption{Model I : Schematic of the LMC plane with various gas flows. The LMC disk is shown with the sense of rotation marked in it. The green arrows show the direction of movement of the LMC centre of mass (at a PA of $\sim$ 80$^\circ$) and the expected direction of the gas stripped due to ram pressure towards NW. The red hexagons show the locations of high velocity gas leaving the LMC as observed in figures 10 and 11. The labels are given analogous to a1, a2 and a3 in figure 11. The location of the SMC is shown. The black arrows shows the possible direction of accreting gas from the MB/SMC. The figure is in the plane of the LMC and the gas is expected to fall into the LMC from behind in a grazing angle.}
    \end{figure}
\begin{figure}
  \resizebox{\hsize}{!}{\includegraphics{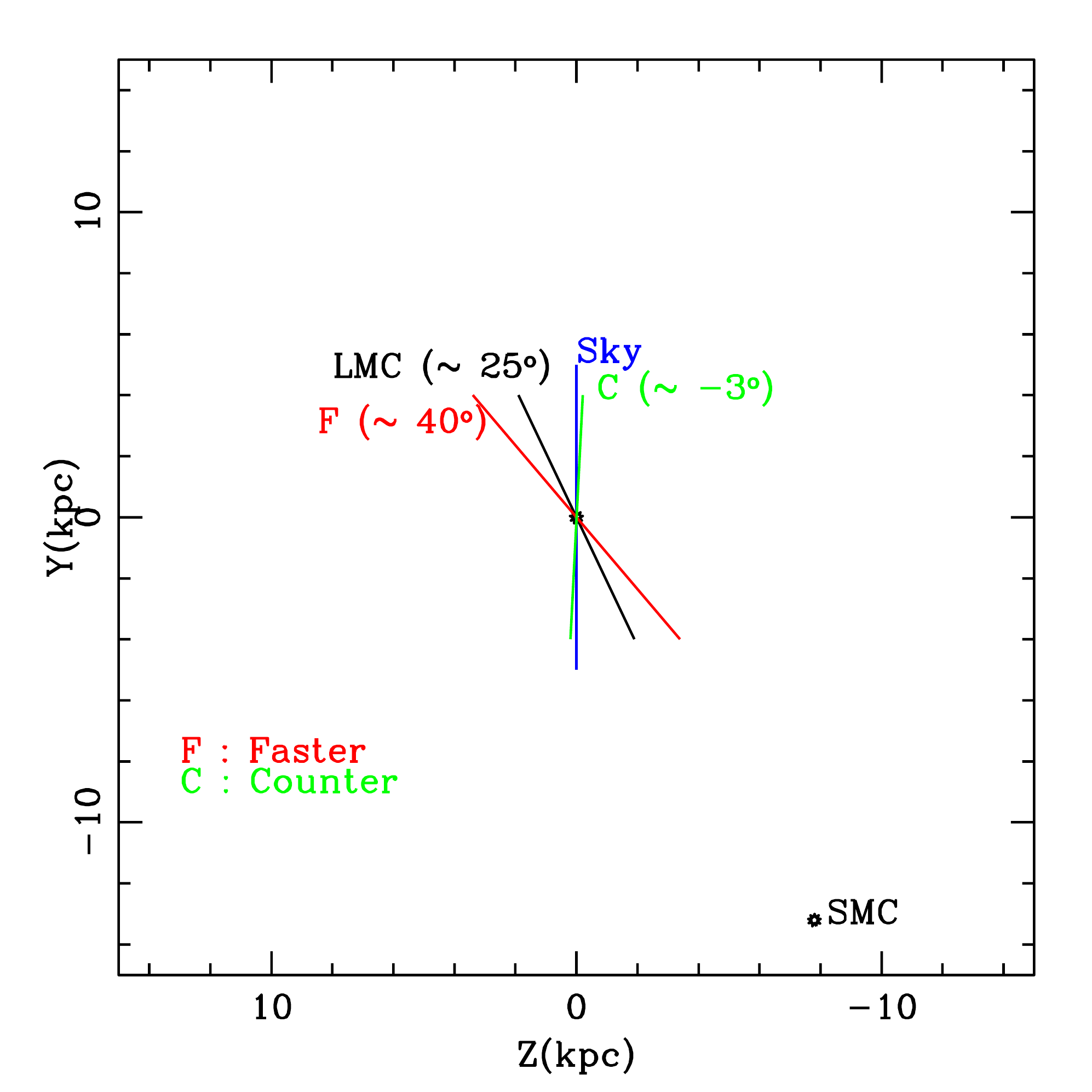}}
   \caption{Model II : Schematic of the edge-on view of the LMC plane and the estimated planes of gas accretion. The plane of the sky and the position of the centre of the SMC are also shown. The angular deviations (inclinations) of each of the planes from the sky plane are marked.}
    \end{figure}

\subsection{Scenario II : Outlier gas in planes shifted from the disk.}

\hspace{4ex}Another scenario to explain the outliers is when it is present in different planes with respect to the LMC disk. The gas can appear to be type 1 if it actually belongs to another plane that is inclined in the opposite sense as the LMC plane. Similarly, the fast type 2 component can appear faster with respect to the LMC disk if the gas is located in a plane with higher inclination. Also, the slow type 2 component can appear slower, if the gas is located in an almost face-on plane. In this model II, the outliers are modelled to be located in three different planes with respect to the LMC disk. In order to derive the parameters of these three planes, one would ideally like to perform a complete fitting of the outlier component, but this does not converge to estimate the parameters due to less spatial coverage of data points. Hence we tried to estimate the approximate values of the parameters of the planes, which closely reproduce the velocity features of the outliers. The estimated planes are shown in figure 14, in ZY coordinates. We identify that there is a nearly face-on plane, shown in figure 14 as disk C (with $i = 0.0$ to $-3^{\circ}.0$), where the type 1 gas is located. The type 1 Arm E would then belong to a plane, which is inclined $28^{\circ}$ away from the plane of the LMC. The slower type 2 kinematic component has velocity very close to the systemic velocity and hence should be almost face on with $i \sim 4^{\circ}$. Thus, the type 1 gas as well as the slower type 2 component are nearly face on, but their planes are significantly deviant from the LMC disk. This scenario also supports the suggestion that the slow type 2 and type 1 outliers could be connected. In a similar estimate done by \citet{2011ApJ...737...29O} with RSGs, their outlier component similar to type 1 was estimated to belong to a plane $54^{\circ}$ away from the plane of the LMC, that is, with an inclination of $i = -19^{\circ}$ in their estimation.

In the case of Arm B, the likely inclination is found to be about $40^{\circ}$, which is $15^{\circ}$ higher than the plane of the LMC. This is shown as disk F in figure 14. Thus, Arm B can be a type 2 component with expected in-plane velocity, but appears faster due to its different inclination. Thus, the analysis supports the possibility of outliers belonging to planes with significantly different inclination, separated by a maximum angle of about 43$^\circ$. A close look at figure 14 shows that these different planes can cause significant changes in the structural parameters of the LMC disk. Since Arm B is expected to be in the plane F and connected to the MB, the model suggests that both Arm B and the MB are behind the disk of the LMC. While the orientation of plane C suggest that Arm E is located in front of the LMC disk. Scenario II is unable to predict the possibility of gas flows along the outlier features, such as Arm B or Arm E, because the velocity deviations observed here are not true deviations in velocity but the result of viewing perspective/projection effect. As the deviations in the velocity are due to viewing perspective, the gas is assumed to have the same velocity as the LMC disk at any given location, thus the question of infall or outflow does not arise. The suggestion that the high velocity streams that we identify in the west, including Arm W and Arm S, are probably caused by the motion of the LMC in the MW halo, is valid according to this scenario also. The presence of high velocity gas identified in the west of the LMC disk as well as in the south and SW of the MB, which cause outflows from the LMC due to ram pressure effects also comes well within this scenario.

Thus, we have put forward two scenarios to explain the kinematic outliers present in the gas. The two scenarios differ in the explanation of the most debated arms, Arm B and Arm E, but have similar explanation for the high velocity gas creating a possible outflow and the arms W and S. Both models explain various observed features in the LMC disk. According to scenario I, the outliers are in the plane of the LMC, while according to scenario II they are in different inclinations. Scenario I predicts the maximum range of velocity assuming all the gas to be in-plane and scenario II predicts the maximum inclination possible to explain the observed velocity. 
It is quite possible that the real situation is a combination of the above two models. That is, the gas could be located in disks with lesser inclination and moving with faster/slower velocity. 

\section{Results and discussion}
\hspace{4ex}We have modelled the H {\sc i} disk of the LMC after correcting the observed velocity field for the transverse motion of the LMC using the recent proper motion measurements by \citealt{2008AJ....135.1024P} and \citealt{2013ApJ...764..161K}.  Corrections are applied for the precession and nutation motion of the LMC disk as well. The data used for the analysis is the intensity weighted mean velocity maps from two data sets. The modelled H {\sc i} velocity distribution, estimated using ring analysis after removing the outliers, successfully reproduces the observed velocity field. The value of PA of kinematic major axis we derived, $126^\circ\pm23^\circ$ is comparable with the value estimated from the stars ($142^\circ\pm5^\circ$). Therefore, the discrepancy  between gas and stellar kinematics in the LMC disk is effectively removed. The modelled H {\sc i} disk estimated using the above mentioned two proper motion values were found to be similar. The effect of precession and nutation of the disk on the estimation of PA and circular velocity is also analysed. We traced the kinematically distinct features present in the LMC. Most of the H {\sc i} gas in the LMC ($\sim 87.9\%$ of the data points) is located in the disk and follows the disk kinematics, which gives the confidence that the estimated modelled disk is close to the true disk of the LMC. The mean H {\sc i} disk shows signs of disturbance within the central 1$^o$.0 degree, might be due to the presence of the bar. The disk region between 1.$^o$0 - 2.$^o$9 is found to be relatively undisturbed. The disk outside 2.$^o$9 is found to be disturbed, likely to be due to the motion of the LMC in the MW halo and related ram pressure. Using ATCA/Parkes data, we detect 12.1$\%$ of the data points as kinematic outliers. We identify 2.7$\%$ of the total data to be type 1, and 3.9$\%$ to be slow type 2. A significant part of these two components is identified with Arm E. We found 5.5$\%$ of the total sample to be fast type 2. We identified the well-known Arm S, Arm W, Arm B and a new stream, Outer Arm, as part of this fast type 2 outlier component. 

From the GASS data analysis, we find evidence for the connection of the LMC to the MB through Arm B and possibly through the Outer Arm. There could be on-going gas accretion to the LMC from the MB through Arm B, if the outlier gas is in-plane with the LMC. 
There is also a suggestive acceleration when it falls into the LMC. This is visualised in figures 11 and 12 by the light blue points at the end of Arm B towards the MB and deep blue points near the 30 Dor region. This suggestion of acceleration is valid only if the gas is located near the LMC plane. Gas located in the inner part of the region connecting Arm B to MB has negative velocity (figures 11, 12). On the other hand, if we go farther away, beyond Y $\sim$$-$6, the velocity of the gas becomes progressively positive, suggesting that gas from this part could leave. This can be clearly seen in figure 11, which shows that there is a velocity gradient along the MB. 
Thus, gas located beyond X=$\sim$ 2.0 and Y= $\sim$ $-$8.0 (figures 11, 12), could be leaving the LMC. Thus, we suggest that there is a probable gas infall into the
LMC from the inner part of the MB, and there is a possible outflow from the south and south-western end of the MB.
In summary, the MB is being sheared, probably by the combined effect of  gravitational potential of the LMC (tidal effect) and its motion in the halo of the MW (hydrodynamical effect). Thus, the outflow as suggested by \citet{2008ApJ...679..432N} is also identified here, but from the outer regions of the MB.

According to the two scenarios we have proposed, the gas in arms such as arm W and S, move either 
faster and/or inclined to the disk. According to \citet{2009MNRAS.399.2004M}, the gas in the outer part of the disk can move to the inner LMC due to the motion of the LMC in the MW halo. The NE region is expected to be compressed because of motion of the LMC and the gas located here can get compressed. This compressed gas could move to inner regions with higher velocity. We speculate that this gas could give rise to arm W and probably arm S. Hence, these arms are created because of motion of the LMC in the MW halo and the resultant inward motion of the gas. This is thus a hydrodynamical effect caused by the motion of the LMC. 

The Arm E is another most debated H {\sc i} feature in connection with gas infall/outflow from the LMC.  The connection of LA with Arm E, as well as the other connected features like the LAI, LAII etc., that are closer to us, make us believe that the Arm E feature is probably in front of the LMC disk.
We find that the Arm E has two kinematically distinct components, a slow type 2 component and a type 1 component. The co-location of these two components lead us to suggest that they belong to the same Arm E and that both the outliers have similar origin. The type 1 gas could be considered as counter-rotating if it is in the plane of the disk. On the other hand, connection of Arm E to LA suggest that it could be in front of the disk, which is consistent with our scenario II, where Arm E is located in an almost face on plane. Thus, Arm E is either counter rotating or in front of the disk (or a combination of both), which can make the gas leave the LMC plane from the SW region. Thus, Arm E is likely to be either stationary or create an outflow to LA.  
In our scenario I, we had suggested that a gas infall from the MB in the western side could create type 1 gas. \citet{2012ApJ...750...36D} had found an accreted component in the LMC, which appears as a polar ring. They attribute this to the earlier interaction of the LMC with the SMC at 1.92 Gyr ago and forms a feature resembling a polar ring about 0.68 Gyr ago. The almost face on feature of the type 1 and slow type 2 outliers appear to have similar orbit and could be the relic of such an interaction. If this is true, then gas in these outliers originated in the SMC.
In the rotation curve, shown in figure 3, panel b, we notice a kink at a radius of $\sim$ 3.$^o$0. This kink could be produced by the presence of the slower/counter-rotating H {\sc i} gas present in the Arm E, which is also located at a similar radius.  

A filament of gas was detected by \citet{2008ApJ...679..432N} emerging from a location close to the southern SGS, which was associated with Arm B and was traced out of the LMC into the MS. This was also traced in the ATCA/Parkes data (figure 12, \citealt{2008ApJ...679..432N}). \citet{2008ApJ...679..432N} concluded that the outflows are being expelled by the supernovae and high stellar winds coming from the SGS. The motion of the LMC through the MW halo is expected to create ram pressure stripping of gas from the LMC disk. This is probably what we notice from the western, southern and SW part of the disk. We propose that these outflows are in general caused by the ram pressure stripping, suggested by the distribution of high velocity gas along the periphery of the MB. We also detect a few outflows on the western side of the LMC disk, probably reaching the MS. The outflow from the MB could be a combination of tidal and hydrodynamical effect, whereas the outflow from the LMC disk is likely to be purely hydrodynamical. We note that
the majority of the faster outliers traced in this analysis are located outside 3.$^o$0 radius. This rising trend in the rotation curve is also noticed after the kink at this radius. Thus, the rise observed in the modelled rotation curve could be caused by the faster gas in the outer regions cause by hydrodynamical compression and ram pressure. Thus, the rising H {\sc i} rotation curve beyond 3.$^o$0 may just be an effect due to the motion of the LMC in the MW halo.

The most important observation from our analysis is that the MB is being sheared.  The gas closer to the LMC disk has the possibility of getting accreted to the LMC and the gas located further away could leave the LMC. Hence, the MB is identified as a possible source of accretion as well as outflow. The accretion through the MB (which in turn is connected to the SMC) is a tidal feature that seems to be modified by the motion of the LMC in the MW halo. 
In summary, we propose that there is a possible ongoing gas infall from the MB though Arm B and Outer Arm. Arm E could either be a stationary feature or a source of outflow. Arms W and S are caused by the motion of LMC in the MW halo. Possible outflows from the west, SW and southern part of MB and LMC are detected, probably due to ram pressure. Thus, the outliers identified here are likely to be the by-products of tidal and hydrodynamical effects in the LMC disk. 
This study shows that this nearest pair of interacting galaxies is ideal to understand the nuances of interactions and gas accretion/outflows. As these details are not available in far away interacting pairs, this is an ideal test bed to understand the tidal and hydrodynamical effects on the H {\sc i} gas. 
\section{Conclusions}
\begin{enumerate}
\item The proper motion correction has a significant role in estimating the kinematics of the LMC H {\sc i} disk. The precession and nutation correction bears a small, but inevitable effect on the estimation of kinematic parameters. Both recent proper motion estimates (\citealt{2008AJ....135.1024P} and \citealt{2013ApJ...764..161K}) were found to produce similar disk parameters, for a given value of $di/dt$.
\item Applying the recent proper motion estimates to the ATCA/Parkes data, the PA of kinematic major axis of the H {\sc i} disk of the LMC (126$^\circ$ $\pm$ 23$^\circ$) is found to be consistent with that of the stellar distribution (142$^\circ$ $\pm$ 5$^\circ$, \citealt{2011ApJ...737...29O}).
\item The modelled H {\sc i} disk shows signs of disturbance within the central 1$^o$.0 degree, might be due to the presence of the bar. The disk region between 1.$^o$0 - 2.$^o$9 is found to be undisturbed. Disk outside 2.$^o$9 is found to be disturbed, likely due to the motion of the LMC in the MW halo and related ram pressure effects. 
\item 
Using ATCA/Parkes data, Arm E is identified with type 1 as well as slow type 2 outliers. We identified the well-known Arm S, Arm W, Arm B and a new stream, Outer Arm, as part of the fast type 2 outlier component.
\item The intensity weighted H {\sc i} velocity map using GASS data clearly shows the extension of the LMC disk to the MB, as well as the large extent of the MB. The kinematic analysis also find a velocity gradient in the MB.
\item We proposed two models (in-plane and out-of-plane) to explain the
presence of the kinematic outliers. We find that these models either separately or in combination can explain most of the observed features of the LMC disk. 
\item We suggest that Arm B is probably an infall feature, originating from the inner MB. Arm E could be an outflow feature. We detect possible outflow from the western LMC disk and south and south-western MB. We suggest that the MB is a source for accretion as well as outflow and is being sheared. 
\item We suggest that the various outliers identified in this study are caused by a combination of the hydrodynamical effect, due to the motion of the LMC in the MW halo, and tidal effects. Thus, this nearest pair of interacting galaxies is a gold mine to understand the nuances of interactions and their effect on gas. 
\end{enumerate}
\appendix
\section{Conversion of Coordinates}
\hspace{4ex}The GASS data set is available in equatorial coordinates. The pixel conversions are as follows.
\begin{equation}
\alpha = (y - 126)\,-0.2236 + 79.4,
\end{equation}
\begin{equation}
\delta = (x - 988.92)\,0.08 + 0.00, 
\end{equation}
\hspace{7 cm}and
\begin{equation}
V_{lsr} = (z-1)\,0.8245-468.32.
\end{equation}
Also, the velocity $V_{lsr}$ with respect to Local Standard of Rest (LSR) must be converted to helio centric frame, if we have to compare with our studies on ATCA/Parkes data. The space velocity of the Sun with respect to the LSR is 18.044 km s$^{-1}$ \\

\begin{equation}
V_{los} = V_{lsr} - V_{solar}.
\end{equation}  
where $V_{los}$ is the observed line-of-sight velocity in heliocentric frame, and $V_{solar}$ is the component of Sun's velocity towards the LMC, which is calculated as follows, 
\begin{equation}
V_{solar} = V_{\odot}\, (\cos b \cos b_{\odot} \cos (l-l_{\odot}) + \sin b \sin b_{\odot}).
\end{equation}
where $V_{\odot}$ is the sun's space velocity and $l_{\odot}$,$b_{\odot}$ represents the direction of its motion with respect to the LSR. 
      
\begin{acknowledgements}
                    We are grateful to the anonymous referee for valuable comments, which improved the analysis and the presentation of the results. We gratefully acknowledge Knut Olsen for providing the ATCA/Parkes H {\sc i} data in a preferred format and for useful suggestions. We are grateful to Smitha Subramanian and C.S.Stalin for the remarks and suggestions, which improved the presentation of the paper. We thank Gurtina Besla for her comments on the paper. Thanks to Arun Surya and Rathnakumar for the valuable support in matlab programming.
\end{acknowledgements}

\bibliographystyle{aa}
\bibliography{report3}

\begin{thebibliography}{37}
\expandafter\ifx\csname natexlab\endcsname\relax\def\natexlab#1{#1}\fi

\bibitem[{{Bekki} \& {Chiba}(2007)}]{2007PASA...24...21B}
{Bekki}, K. \& {Chiba}, M. 2007, \pasa, 24, 21

\bibitem[{{Besla} {et~al.}(2007){Besla}, {Kallivayalil}, {Hernquist},
  {Robertson}, {Cox}, {van der Marel}, \& {Alcock}}]{2007ApJ...668..949B}
{Besla}, G., {Kallivayalil}, N., {Hernquist}, L., {et~al.} 2007, \apj, 668, 949

\bibitem[{{Besla} {et~al.}(2012){Besla}, {Kallivayalil}, {Hernquist}, {van der
  Marel}, {Cox}, \& {Kere{\v s}}}]{2012MNRAS.421.2109B}
{Besla}, G., {Kallivayalil}, N., {Hernquist}, L., {et~al.} 2012, \mnras, 421,
  2109

\bibitem[{{Casetti-Dinescu} {et~al.}(2012){Casetti-Dinescu}, {Vieira},
  {Girard}, \& {van Altena}}]{2012ApJ...753..123C}
{Casetti-Dinescu}, D.~I., {Vieira}, K., {Girard}, T.~M., \& {van Altena}, W.~F.
  2012, \apj, 753, 123

\bibitem[{{Diaz} \& {Bekki}(2012)}]{2012ApJ...750...36D}
{Diaz}, J.~D. \& {Bekki}, K. 2012, \apj, 750, 36

\bibitem[{{Fox} {et~al.}(2013){Fox}, {Richter}, {Wakker}, {Lehner}, {Howk},
  {Ben Bekhti}, {Bland-Hawthorn}, \& {Lucas}}]{2013ApJ...772..110F}
{Fox}, A.~J., {Richter}, P., {Wakker}, B.~P., {et~al.} 2013, \apj, 772, 110

\bibitem[{{Indu} \& {Subramaniam}(2011)}]{2011A&A...535A.115I}
{Indu}, G. \& {Subramaniam}, A. 2011, \aap, 535, A115

\bibitem[{{Jones} {et~al.}(1994){Jones}, {Klemola}, \&
  {Lin}}]{1994AJ....107.1333J}
{Jones}, B.~F., {Klemola}, A.~R., \& {Lin}, D.~N.~C. 1994, \aj, 107, 1333

\bibitem[{{Kalberla} {et~al.}(2010){Kalberla}, {McClure-Griffiths}, {Pisano},
  {Calabretta}, {Ford}, {Lockman}, {Staveley-Smith}, {Kerp}, {Winkel},
  {Murphy}, \& {Newton-McGee}}]{2010A&A...521A..17K}
{Kalberla}, P.~M.~W., {McClure-Griffiths}, N.~M., {Pisano}, D.~J., {et~al.}
  2010, \aap, 521, A17

\bibitem[{{Kallivayalil} {et~al.}(2006{\natexlab{a}}){Kallivayalil}, {van der
  Marel}, \& {Alcock}}]{2006ApJ...652.1213K}
{Kallivayalil}, N., {van der Marel}, R.~P., \& {Alcock}, C. 2006{\natexlab{a}},
  \apj, 652, 1213

\bibitem[{{Kallivayalil} {et~al.}(2006{\natexlab{b}}){Kallivayalil}, {van der
  Marel}, {Alcock}, {Axelrod}, {Cook}, {Drake}, \&
  {Geha}}]{2006ApJ...638..772K}
{Kallivayalil}, N., {van der Marel}, R.~P., {Alcock}, C., {et~al.}
  2006{\natexlab{b}}, \apj, 638, 772

\bibitem[{{Kallivayalil} {et~al.}(2013){Kallivayalil}, {van der Marel},
  {Besla}, {Anderson}, \& {Alcock}}]{2013ApJ...764..161K}
{Kallivayalil}, N., {van der Marel}, R.~P., {Besla}, G., {Anderson}, J., \&
  {Alcock}, C. 2013, \apj, 764, 161

\bibitem[{{Kim} {et~al.}(1999){Kim}, {Dopita}, {Staveley-Smith}, \&
  {Bessell}}]{1999AJ....118.2797K}
{Kim}, S., {Dopita}, M.~A., {Staveley-Smith}, L., \& {Bessell}, M.~S. 1999,
  \aj, 118, 2797

\bibitem[{{Kim} {et~al.}(1998){Kim}, {Staveley-Smith}, {Dopita}, {Freeman},
  {Sault}, {Kesteven}, \& {McConnell}}]{1998ApJ...503..674K}
{Kim}, S., {Staveley-Smith}, L., {Dopita}, M.~A., {et~al.} 1998, \apj, 503, 674

\bibitem[{{Kim} {et~al.}(2003){Kim}, {Staveley-Smith}, {Dopita}, {Sault},
  {Freeman}, {Lee}, \& {Chu}}]{2003ApJS..148..473K}
{Kim}, S., {Staveley-Smith}, L., {Dopita}, M.~A., {et~al.} 2003, \apjs, 148,
  473

\bibitem[{{Mastropietro}(2009)}]{2009IAUS..256..117M}
{Mastropietro}, C. 2009, in IAU Symposium, Vol. 256, IAU Symposium, ed. J.~T.
  {Van Loon} \& J.~M. {Oliveira}, 117--121

\bibitem[{{Mastropietro} {et~al.}(2009){Mastropietro}, {Burkert}, \&
  {Moore}}]{2009MNRAS.399.2004M}
{Mastropietro}, C., {Burkert}, A., \& {Moore}, B. 2009, \mnras, 399, 2004

\bibitem[{{Mastropietro} {et~al.}(2005){Mastropietro}, {Moore}, {Mayer},
  {Wadsley}, \& {Stadel}}]{2005MNRAS.363..509M}
{Mastropietro}, C., {Moore}, B., {Mayer}, L., {Wadsley}, J., \& {Stadel}, J.
  2005, \mnras, 363, 509

\bibitem[{{McClure-Griffiths} {et~al.}(2009){McClure-Griffiths}, {Pisano},
  {Calabretta}, {Ford}, {Lockman}, {Staveley-Smith}, {Kalberla}, {Bailin},
  {Dedes}, {Janowiecki}, {Gibson}, {Murphy}, {Nakanishi}, \&
  {Newton-McGee}}]{2009ApJS..181..398M}
{McClure-Griffiths}, N.~M., {Pisano}, D.~J., {Calabretta}, M.~R., {et~al.}
  2009, \apjs, 181, 398

\bibitem[{{Nidever} {et~al.}(2008){Nidever}, {Majewski}, \&
  {Burton}}]{2008ApJ...679..432N}
{Nidever}, D.~L., {Majewski}, S.~R., \& {Burton}, W.~B. 2008, \apj, 679, 432

\bibitem[{{Nidever} {et~al.}(2013){Nidever}, {Monachesi}, {Bell}, {Majewski},
  {Mu{\~n}oz}, \& {Beaton}}]{2013ApJ...779..145N}
{Nidever}, D.~L., {Monachesi}, A., {Bell}, E.~F., {et~al.} 2013, \apj, 779, 145

\bibitem[{{Olsen} \& {Massey}(2007)}]{2007ApJ...656L..61O}
{Olsen}, K.~A.~G. \& {Massey}, P. 2007, \apjl, 656, L61

\bibitem[{{Olsen} {et~al.}(2011){Olsen}, {Zaritsky}, {Blum}, {Boyer}, \&
  {Gordon}}]{2011ApJ...737...29O}
{Olsen}, K.~A.~G., {Zaritsky}, D., {Blum}, R.~D., {Boyer}, M.~L., \& {Gordon},
  K.~D. 2011, \apj, 737, 29

\bibitem[{{Piatek} {et~al.}(2008){Piatek}, {Pryor}, \&
  {Olszewski}}]{2008AJ....135.1024P}
{Piatek}, S., {Pryor}, C., \& {Olszewski}, E.~W. 2008, \aj, 135, 1024

\bibitem[{{Richter} {et~al.}(2013){Richter}, {Fox}, {Wakker}, {Lehner}, {Howk},
  {Bland-Hawthorn}, {Ben Bekhti}, \& {Fechner}}]{2013ApJ...772..111R}
{Richter}, P., {Fox}, A.~J., {Wakker}, B.~P., {et~al.} 2013, \apj, 772, 111

\bibitem[{{Staveley-Smith} {et~al.}(2003){Staveley-Smith}, {Kim}, {Calabretta},
  {Haynes}, \& {Kesteven}}]{2003MNRAS.339...87S}
{Staveley-Smith}, L., {Kim}, S., {Calabretta}, M.~R., {Haynes}, R.~F., \&
  {Kesteven}, M.~J. 2003, \mnras, 339, 87

\bibitem[{{Subramaniam} \& {Prabhu}(2005)}]{2005ApJ...625L..47S}
{Subramaniam}, A. \& {Prabhu}, T.~P. 2005, \apjl, 625, L47

\bibitem[{{Subramaniam} \& {Subramanian}(2009)}]{2009ApJ...703L..37S}
{Subramaniam}, A. \& {Subramanian}, S. 2009, \apjl, 703, L37

\bibitem[{{Subramanian} \& {Subramaniam}(2010)}]{2010A&A...520A..24S}
{Subramanian}, S. \& {Subramaniam}, A. 2010, \aap, 520, A24

\bibitem[{{Subramanian} \& {Subramaniam}(2013)}]{2013A&A...552A.144S}
{Subramanian}, S. \& {Subramaniam}, A. 2013, \aap, 552, A144

\bibitem[{{van der Marel}(2001)}]{2001AJ....122.1827V}
{van der Marel}, R.~P. 2001, \aj, 122, 1827

\bibitem[{{van der Marel} {et~al.}(2002){van der Marel}, {Alves}, {Hardy}, \&
  {Suntzeff}}]{2002AJ....124.2639V}
{van der Marel}, R.~P., {Alves}, D.~R., {Hardy}, E., \& {Suntzeff}, N.~B. 2002,
  \aj, 124, 2639

\bibitem[{{van der Marel} \& {Cioni}(2001)}]{2001AJ....122.1807V}
{van der Marel}, R.~P. \& {Cioni}, M.-R.~L. 2001, \aj, 122, 1807

\bibitem[{{van der Marel} \& {Kallivayalil}(2014)}]{2014ApJ...781..121V}
{van der Marel}, R.~P. \& {Kallivayalil}, N. 2014, \apj, 781, 121

\bibitem[{{van der Marel} {et~al.}(2009){van der Marel}, {Kallivayalil}, \&
  {Besla}}]{2009IAUS..256...81V}
{van der Marel}, R.~P., {Kallivayalil}, N., \& {Besla}, G. 2009, in IAU
  Symposium, Vol. 256, IAU Symposium, ed. J.~T. {van Loon} \& J.~M. {Oliveira},
  81--92

\bibitem[{{van Loon} {et~al.}(2013){van Loon}, {Bailey}, {Tatton}, {Ma{\'{\i}}z
  Apell{\'a}niz}, {Crowther}, {de Koter}, {Evans}, {H{\'e}nault-Brunet},
  {Howarth}, {Richter}, {Sana}, {Sim{\'o}n-D{\'{\i}}az}, {Taylor}, \&
  {Walborn}}]{2013A&A...550A.108V}
{van Loon}, J.~T., {Bailey}, M., {Tatton}, B.~L., {et~al.} 2013, \aap, 550,
  A108

\bibitem[{{Weinberg}(2000)}]{2000ApJ...532..922W}
{Weinberg}, M.~D. 2000, \apj, 532, 922

\end{thebibliography}
\end{document}